\documentclass[preprint,12pt]{elsarticle}

\usepackage{amssymb}
\usepackage{amsmath}
\usepackage{subcaption}
\usepackage{bm}
\usepackage[breaklinks]{hyperref}

\journal{Nuclear Physics B}

\begin{document}

\begin{frontmatter}



% pdflatex XYZ_2.tex; bibtex XYZ_2; pdflatex XYZ_2.tex;

\title{Prospects for Higgs boson and new scalar resonant production searches in $ttbb$ final state at the LHC}


\author[label1]{P.~Mandrik}

\address[label1]{NRC ``Kurchatov Institute'' - IHEP, Protvino, Moscow Region, Russia}
%% \address[label2]{Moscow Institute of Physics and Technology, Dolgoprudny, Moscow Region, Russia}

\begin{abstract}
In this article we probe resonant associated production of a Standard Model Higgs boson with new heavy scalar resonance in proton-proton collisions at a center-of-mass energy $\sqrt{s} = 13$ TeV.
The Higgs boson and new scalar resonant are required to decay into a pair of bottom quarks and a pair of top quarks, respectively. Semileptonic decay of top quarks is considered.
The searches are projected into operation conditions of the Large Hadron Collider during Run II data taking period at a center-of-mass energy of 13 TeV using Monte Carlo generated events, realistic detector response simulation and available Open Data samples.
Analysis strategies are presented and machine learning approach using Deep Neural Network is proposed to resolve ambiguous in jets assignment and improve kinematic reconstruction of signal events.
Sensitivity of the CMS detector is estimated as $95\%$ expected upper limits on the product of the production cross section and the branching fractions of the searched particles.
\end{abstract}


\end{frontmatter}


\section{Introduction} \label{section_intro}

\begin{figure}
  \centering
  \includegraphics[width=0.7\linewidth,clip]{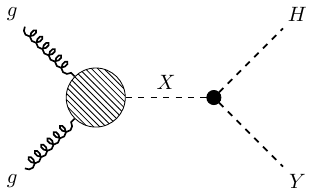}
  \caption{ Feynman diagram of the $gg \rightarrow X \rightarrow Y H$ process. }
  \label{xyh_decays}
\end{figure}

Since the year 2012 when the ATLAS and CMS Collaborations at the Large Hadron Collider (LHC) have discovered a new particle $H$ with a mass of about 125 GeV \cite{ATLAS:2012yve, CMS:2012qbp, ATLAS:2016neq}
the question whether the observed scalar boson forms part of an extended Higgs sector is one of the science drivers for on-going research and studies at future colliders.
Indeed, while the current experimental measurements of the properties of this particle agree with the predictions for the Higgs boson of the Standard Model (SM),
they are also in some cases compatible with the interpretation as a Higgs boson in a variety of SM extensions corresponding to different underlying physics.
Among the beyond the standard model (BSM) theories, which address a number of open fundamental theoretical questions and striking observations in nature,
the minimal supersymmetric SM extension (MSSM) features two charged and three neutral Higgs bosons, one of which can be associated with $H$ \cite{Haber:2017aci}.
The next-to-minimal supersymmetric standard model (NMSSM) introduces one additional complex singlet field to MSSM, resulting in two charged, three neutral scalar and two neutral pseudoscalar Higgs
bosons \cite{Ellwanger:2009dp, Maniatis:2009re}. 
In the NMSSM the more massive Higgs bosons is allowed to asymmetric decay into lighter Higgs bosons, which in the context of the LHC leads to a process (Fig. \ref{xyh_decays}):
\begin{eqnarray}
  pp \, \to \, X \, \to Y H \to SM
\label{rat1}
\end{eqnarray}
where $X$ and $Y$ are new massive scalar resonances, $H$ - SM Higgs boson, and $SM$ stands for SM particles in final state. 

In some scenarios $Y$ could have significant suppression of its couplings to SM particles and thus of its direct production at the LHC \cite{Munir:2013dya, PhysRevD.90.095014}. 
In this case, the production chain (\ref{rat1}) would become the dominant source for $X$ and $Y$ particles.
The same topology arise in the two-real-scalar-singlet model (TRSM) \cite{Robens:2019kga} where two additional singlet fields are added into SM
and mixed into three physical scalar states.
The search of new resonances decaying into pair of Higgs bosons is also motivated by models with warped extra dimensions that predict heavy spin$-0$ radion \cite{Bae:2000pk, Csaki:2000zn}
or the first Kaluza-Klein (KK) excitation of a spin-2 graviton \cite{PhysRevD.76.036006,Oliveira:2014kla} and 
Two-Higgs-Doublet Models (2HDM) with two neutral CP-even scalars, a neutral CP-odd pseudoscalar and two charged Higgs bosons \cite{Branco:2011iw, Davidson:2005cw, Coleppa:2013dya}.

The first search for such signature (\ref{rat1}) at the LHC was presented recently by CMS Collaboration \cite{CMS:2021yci} where $\tau \tau bb$ final state is used. 
However, many other final states are uncovered at the moment. 
One of the most promising decay channel is $Y \rightarrow t\bar{t}$, because of the especial role of top quark in Higgs sector and possibility to exploit a signature 
of the top quarks decays to select and reconstruct events \cite{Boos:2019ffk, Dawson:2002tg, Cepeda:2019klc}.
The top quark is the heaviest of all known elementary particles.
For new SM-like Higgs bosons with mass greater than $2 \cdot m_{t}$ the branching fraction of the decay into top quarks pair is enriched in comparison to $b\bar{b}$ channel dominated for SM Higgs (see e.g. \cite{dEnterria:2012eip}).
And as long as scalar boson $Y$ generated by pure singlet extensions of SM, $Br(Y \rightarrow NP) = 0$, no Higgs-to-Higgs decays are possible for $Y$,
the $Y$ has branching fractions identical to a SM-like Higgs boson of the same mass.
Some benchmark scenarios of TRSM (see Fig. 10 of \cite{Robens:2019kga}) and NMSSM \cite{Ellwanger:2022jtd} promote $b\bar{b}t\bar{t}$ for heavy $Y$ as one of the prominent $YH$ decay channel.
The another notable Higgs decay to two photon in SM is mediated by triangular loops of charged fermions as well as massive vector boson and driven by interaction strength of Higgs with top quarks.
The anomalous interactions of Higgs bosons with top quarks are less constrained than with light quarks by various low-energy precision measurement \cite{Harnik:2012pb, Ilyushin:2019mkp}. 
While $H$ boson production in association with a top quark-antiquark pair is actively investigated at the LHC based on Run I and Run II data-taking eras,
the analyzes are focused on non-resonant low energy kinematic regions.
The observation of $t\bar{t}H$ production was reported for $H$ decays to pairs of $W$ bosons, $Z$ bosons, photons, tau leptons, or bottom quark jets \cite{ATLAS:2018mme, CMS:2018uxb}.
Measurement for the $t\bar{t}H$ together with $tH$ SM processes is done by CMS Collaboration at $\sqrt{s} = 13$ TeV in final states with electrons, muons, and hadronically decaying tau leptons \cite{CMS:2020mpn}. 
The reported production rates for the $t\bar{t}H$ and $tH$ signals are within uncertainties with of their standard model (SM) expectations.
Moreover, the final state with pair of two top quarks is uncovered by resonant and non-resonant di-Higgs production searches at the LHC \cite{CMS:2018ipl, Guerrero:2021dtw, Mandrik:2019pis, Veatch:2022uzz}.

In this article we study the new X resonance production with subsequent decay of X into new Higgs like particle Y and SM Higgs at the LHC conditions: 
\begin{eqnarray}
  pp \, \to \, X \, \to Y H \, , \, Y \to t \bar{t}  \, , \, H \to b\bar{b}
\label{rat2}
\end{eqnarray}
where semileptonic decay of top quarks ($t \bar{t} \rightarrow b \bar{b} q \bar{q}' \ell^{\pm} \nu$) is considered as most sensitive \cite{CMS:2018hnq, CMS:2018sah}. Section \ref{section_a} covers the generation of simulated events used to describe signal and dominated background processes.
Section \ref{section_b1} describes the analysis of parton-level distributions over kinematic variables, 
while the study of events after the detector reconstruction and cut-and-count analysis is given in Section \ref{section_b2}.
The presence of four jets from $b$-quarks and two light jets in final state make it challenging to perform the reconstruction the event kinematic. 
Indeed, we use an advantage of Deep Learning techniques for the signal kinematic reconstruction and event selection, discussed in Section \ref{section_b2}.
We end the paper with results of statistical inference in Section \ref{section_b4} and a brief summary in Section \ref{section_c}.

\section{ Event simulation }  \label{section_a}
NMSSMHET model \cite{Curtin:2013fra} is used to generate signal production from gluon-gluon fusion.
We consider $X$ and $Y$ bosons to be narrow scalars resonances (decay width set to 1 MeV) with branching ratios $B(X \rightarrow YH)$ and $B(Y \rightarrow t\bar{t})$ set to 100\%.
The model is interfaced with {\scshape MG5\_}a{\scshape MC@NLO} 2.7.3~\cite{Alwall:2014hca} package at LO precision using the UFO module~\cite{Degrande:2011ua}.
The decays of $Y$ boson, top quarks and $W$-bosons are performed using MadSpin \cite{Artoisenet:2012st} package to decrease CPU cost of event Monte Carlo (MC) generation while preserving spin correlation effects.
The signal generation is performed for the mass ranges of $650 \leq m_{X} \leq 1900$ GeV and $375 \leq m_{Y} \leq 1600$ GeV.
Samples are produced for both cases when either top quark or $\bar{t}$-quark decay into leptons with electron or muon in final state.
All generated events are processed with {\scshape Pythia}~8.306~\cite{Sjostrand:2014zea} for showering, hadronization and the underlying event description.
The {\scshape NNPDF3.0} \cite{Ball:2014uwa} parton distribution functions set is used.
The detector simulation has been performed with the fast simulation tool~{\scshape Delphes}~3.5.0~\cite{deFavereau2014} using the CMS detector~\cite{CMS:2008xjf, cms_git_link} parameterization cards. 
No additional pileup interactions are added to the simulation.

For the backgrounds the released under the Creative Commons CC0 waiver \cite{CC0} Open Data samples are used \cite{ttbar_data, ttH} with events available after detailed detector simulation based on GEANT 4~\cite{geant4} to model experimental effects, such as reconstruction, selection efficiencies, and resolutions in the CMS detector.
The samples are corresponded to LHC CMS Run II 2015 collision data. The $t\bar{t}+0,1,2$ jets and irreducible SM $t\bar{t}H+0,1$ jets backgrounds are generated with {\scshape MG5\_}a{\scshape MC@NLO} at NLO and
interfaced with {\scshape Pythia} 8 using FxFx merging scheme for the parton showering \cite{Frederix:2012ps}.
The $t\bar{t}+$jets sample is further separated into the following processes based on the flavour of additional jets that do not originate from the top quark decays in the event:
$t\bar{t}+$heavy flavour jets ($t\bar{t}+$hf) defined at generator level as the events in which at least one additional $b$ or $c$ jet is generated; 
$t\bar{t}+$light flavour jets ($t\bar{t}+$lf) which corresponds to events that do not belong to $t\bar{t}+$hf group.
% Single top quark $t-$channel and $Wt-$channel samples are generated at NLO and matched with parton showers using {\scshape POWHEG} method \cite{Frixione:2007vw, Re:2010bp, Alioli:2009je}.

\section{ Event analysis }  \label{section_b}
\subsection{ Parton-level }  \label{section_b1}
\begin{figure}
  \centering
  \includegraphics[width=0.50\columnwidth]{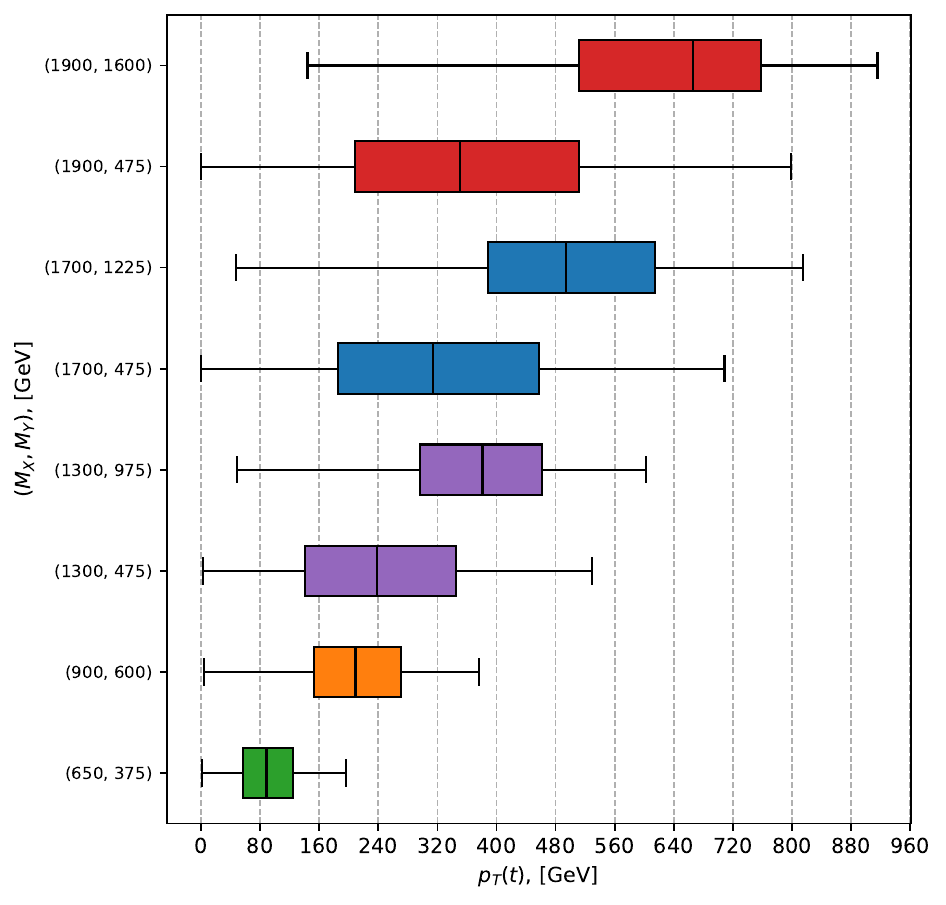}
  \includegraphics[width=0.46\columnwidth]{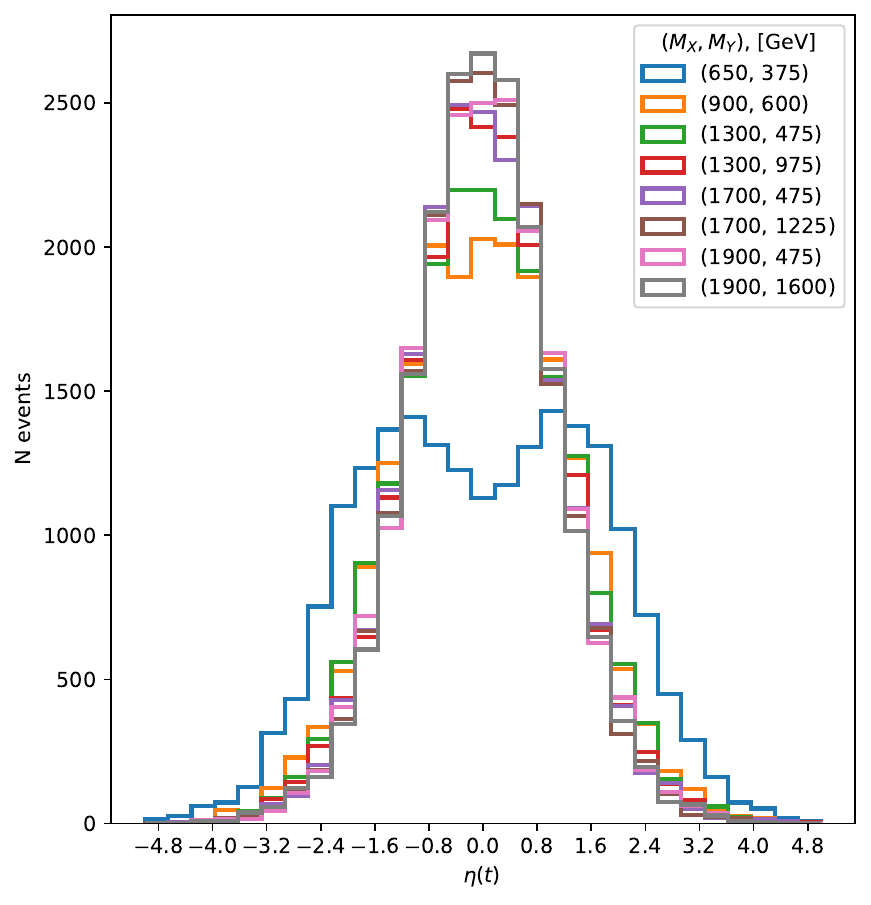} \\
  \caption{ left: box plots\protect\footnotemark for transverse-momentum distributions of top quarks from $Y$ decay. Right: pseudorapidity distributions of top quarks. Results are given for different mass scenarios of $X$ and $Y$ resonances. }
  \label{fig:fcc_top}
\end{figure}

\footnotetext{Box plot is defined by minimum, the maximum, the sample median, and the first and third quartiles. Median is the middle value in the data set. First quartile $Q_1$ (left box edge) is the median of the lower half of the dataset. Third quartile $Q_3$ (right box edge) is the median of the upper half of the dataset. Minimum (left whisker) is defined as $Q_1 - 1.5 \times (Q_3 - Q_1)$. Maximum (right whisker) is defined as $Q_3 + 1.5 \times (Q_3 - Q_1)$. See \cite{wiki_box_plot} for reference.}

The final state consists of a pair of $b$ jets from $H$ decay, a two $b$ jets from top quarks pair decays, a pair of light jets from hadronic decay of $W$ boson, charged lepton and transverse component of the  momentum $p_T^{miss}$ of neutrino from leptonic decay of $W$ boson. 

\begin{figure}
  \centering
  \includegraphics[width=0.95\columnwidth]{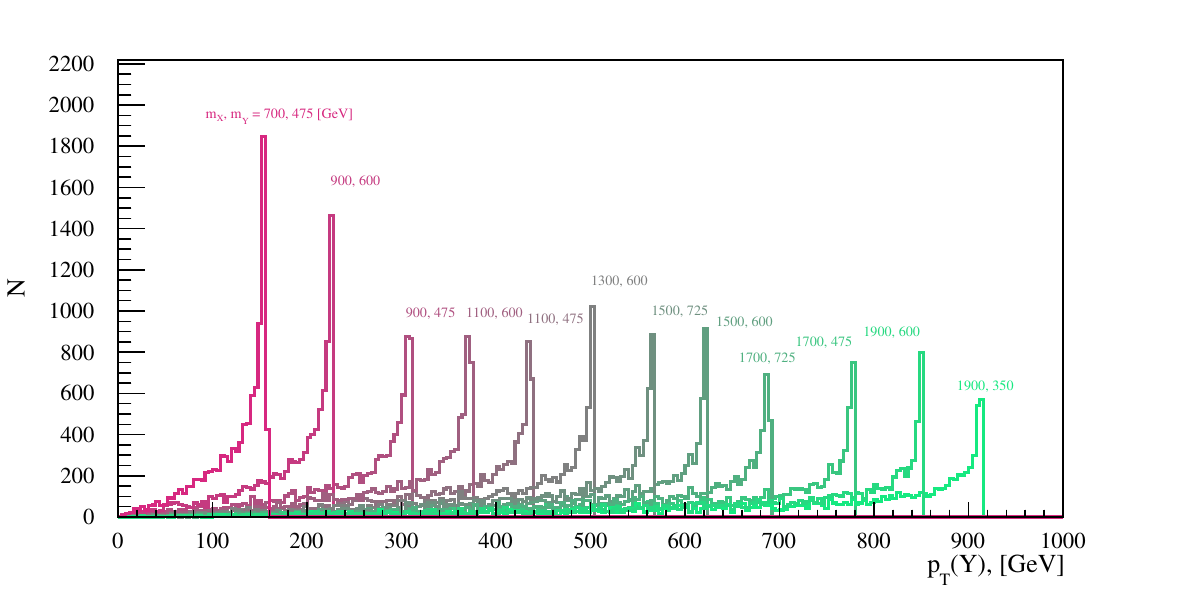}
  \caption{ Transverse-momentum distributions of $Y$ resonances (same for $H$ at parton-level at generation step) for different mass scenarios of $X$ and $Y$. }
  \label{pt_Y}
\end{figure}

Top-quarks from $Y$ decays are populated regions approximately from $80$ GeV to $800$ GeV (Fig. \ref{fig:fcc_top}) for considered mass scenarios. 
The analysis sensitivity may be enriched by focusing on easier to reconstruct and distinguish so-called highly boosted topology of top quarks (e.g. \cite{, CMS:2020ttz, Mandrik:2018yhe, CMS:2019fak, ATLAS:2018orx}) as the top quark decay products are collimated into a large-radius jet by the Lorentz boost of the top quarks. But the measurements of boosted top quarks at the LHC are performed only starting from $p_{T} > 400$ GeV \cite{CMS:2019fak} and $p_{T} > 500$ GeV \cite{ATLAS:2018orx}. The boosted regime of top quarks is also not dominated up to $1000$ GeV \cite{pmandrik} and sufficient fraction of events has a clear resolved semileptonic signal signature of top quarks pair decay, on which we will focus in the following analysis. 
Full kinematic reconstruction of resolved events will give a clear possibility to detect signal process as peak in $X$ and $Y$ invariant mass distributions as well as to probe other sensitive variables (such as transverse-momentum of $Y$ resonance shown at Fig. \ref{pt_Y}). 
Essential step for this is to unravel $b$ quarks origins.

\begin{figure}
  \centering
  \includegraphics[width=0.47\columnwidth]{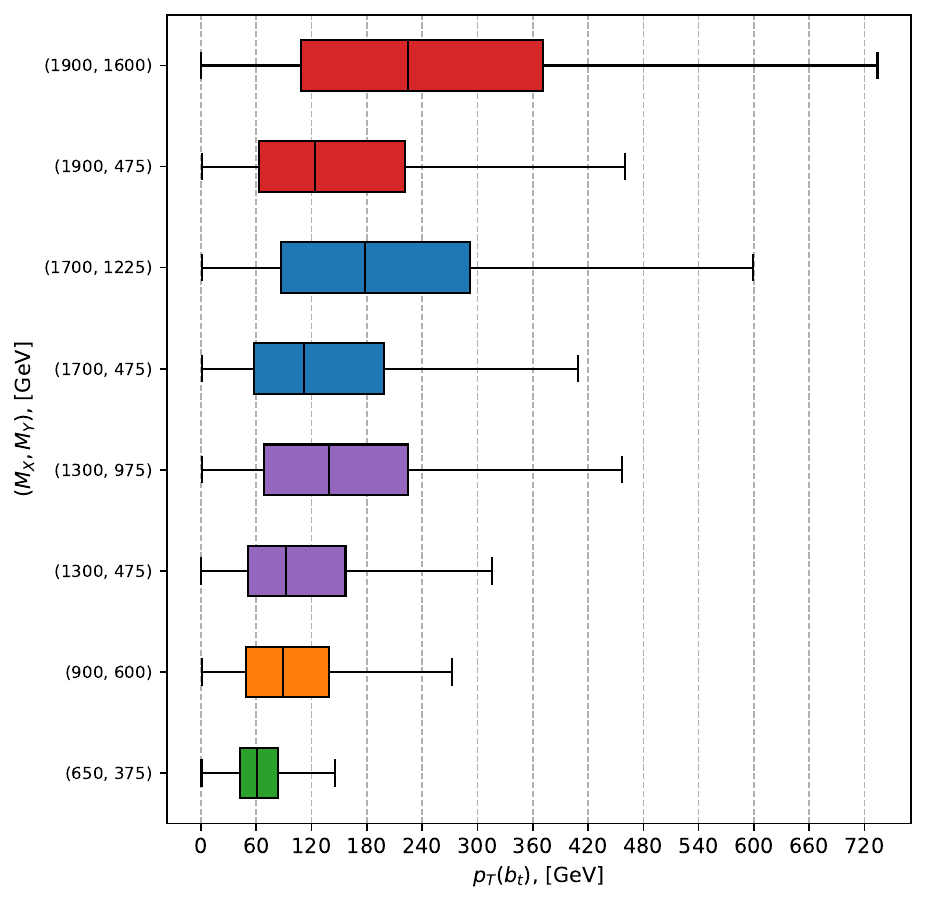}
  \includegraphics[width=0.47\columnwidth]{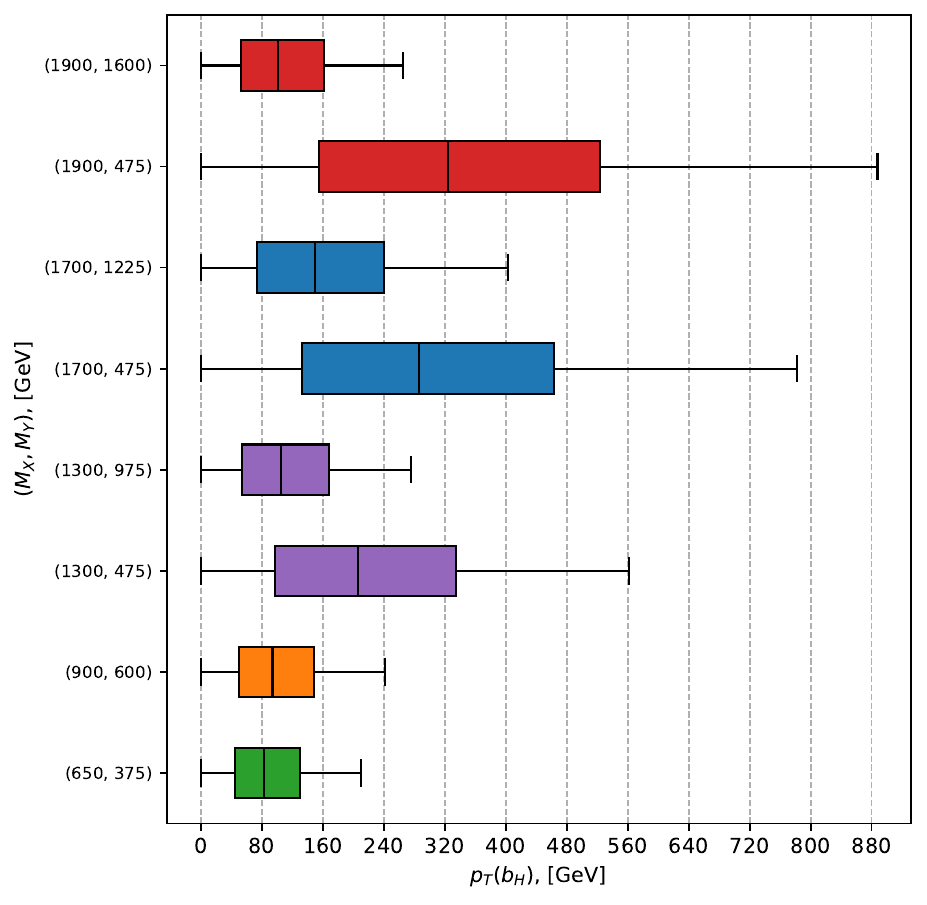} \\
  \caption{ box plots for transverse-momentum distributions of $b$-quarks from top quarks pair decays (left) and from $H$ decay (right). Results are given for different mass scenarios of $X$ and $Y$ resonances. }
  \label{fig:fcc_bot}
\end{figure}

\begin{figure}
  \centering
  \includegraphics[width=0.47\columnwidth]{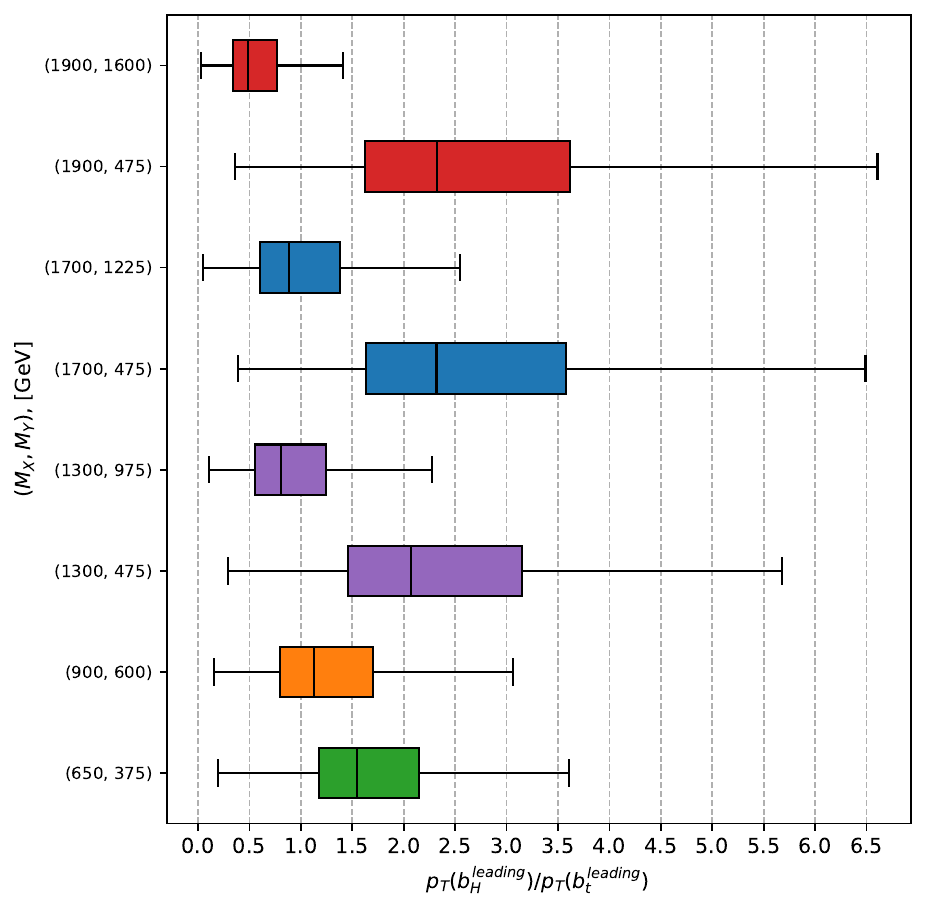}
  \includegraphics[width=0.47\columnwidth]{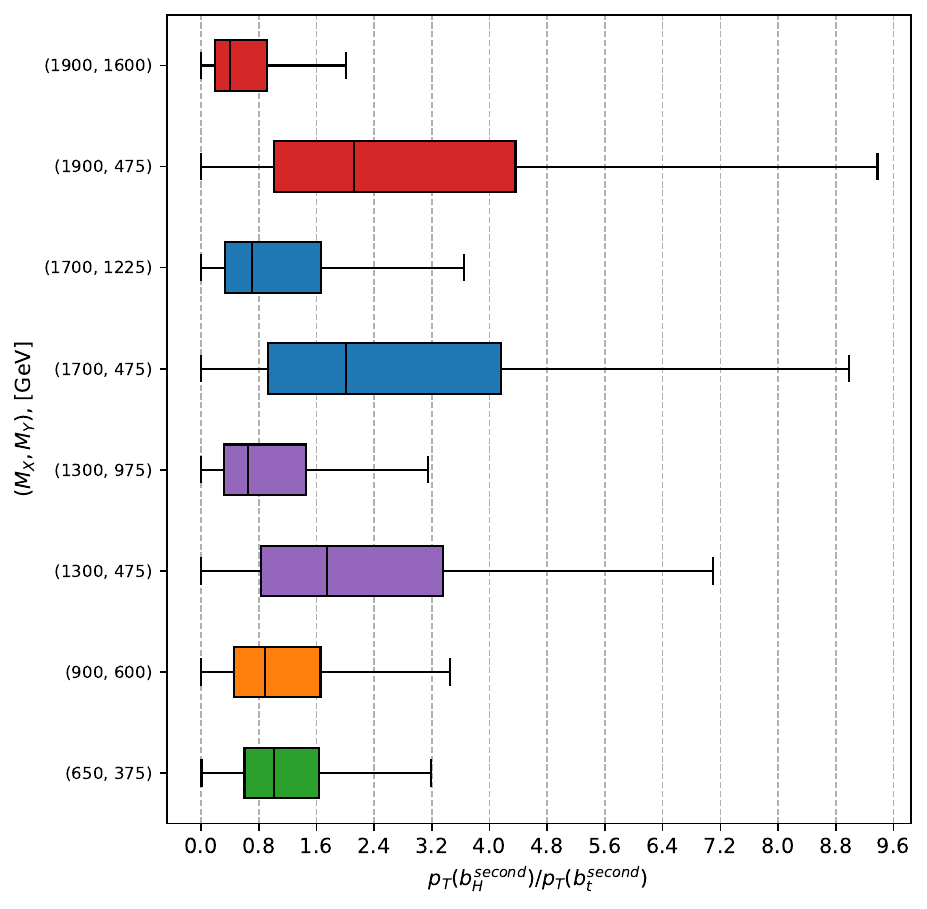} \\
  \caption{ box plots for transverse-momentum ratio distributions of $b$-quarks $p_T$ from $H$ decay to $b$-quarks $p_T$ from top quarks pair decays. Leading (left) and subleading (right) in $p_T$ quarks are used. Results are given for different mass scenarios of $X$ and $Y$ resonances. }
  \label{fig:fcc_bot2}
\end{figure}

For the scenarios with light mass of $Y$ boson the $b_H$ quarks from $H$ decay tend to be more energetic than $b_t$ quarks from top quarks pair decays, whose $p_T$ is rising with the increase of $Y$ mass (Fig. \ref{fig:fcc_bot}). 
Box plots for transverse-momentum ratio distributions of $b$-quarks $p_T$ from $H$ decay to $b$-quarks $p_T$ from top quarks pair decays are given in Fig. \ref{fig:fcc_bot2}.
For several mass scenarios, when the distributions of $p_T (b_H)/p_T (b_t)$ either below or above $1$, this feature can be used as a separation criterion. On the other hand, $p_T$ regions populated by $b_H$ and $b_t$ quarks significantly overlapped for others considered scenarios complicating the task of event reconstruction.

\begin{figure}
  \centering
  \includegraphics[width=0.49\columnwidth]{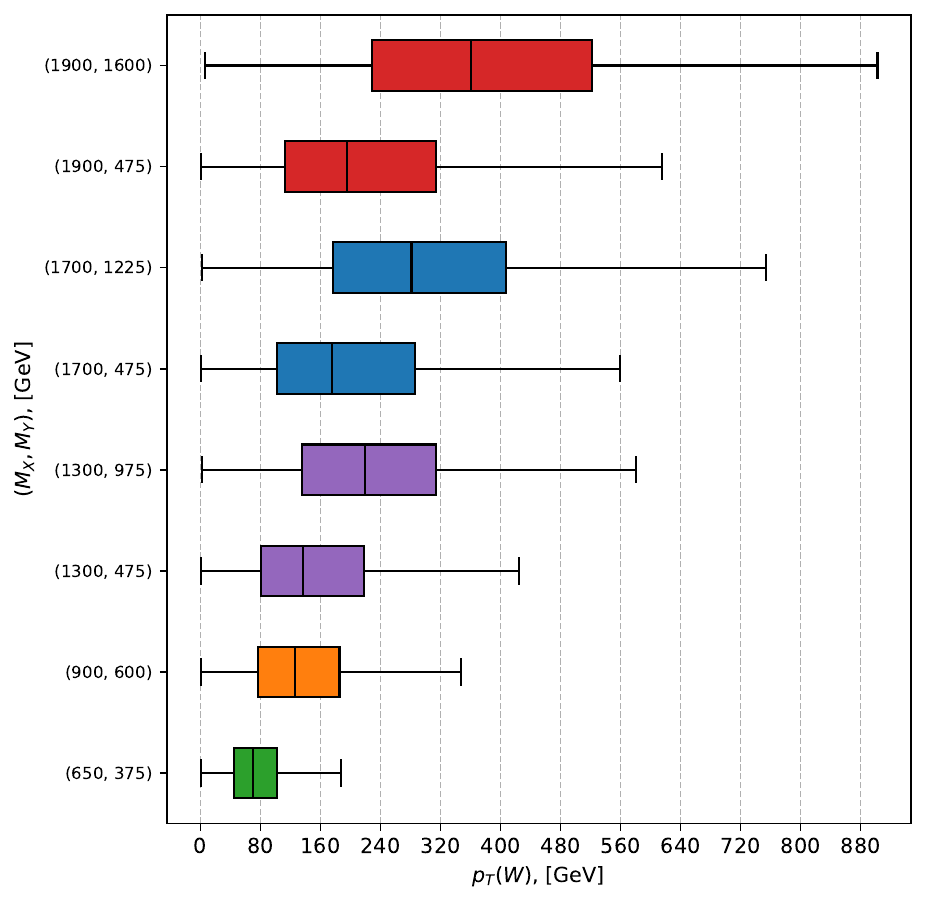}
  \includegraphics[width=0.46\columnwidth]{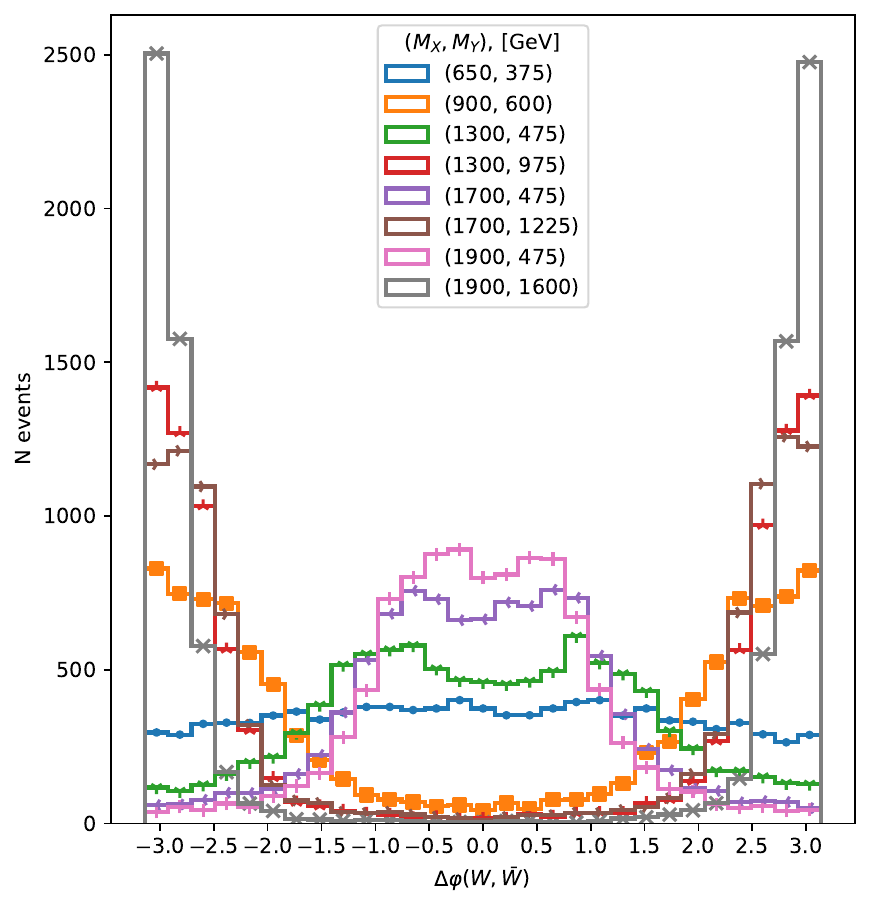} \\
  \caption{ Left: box plots for transverse-momentum distributions of $W$ bosons from top quarks decays. Right: $\Delta \varphi$ between $W$ bosons from top quarks decays. Results are given for different mass scenarios of $X$ and $Y$ resonances. }
  \label{fig:fcc_W}
\end{figure}

\begin{figure}
  \centering
  \includegraphics[width=0.47\columnwidth]{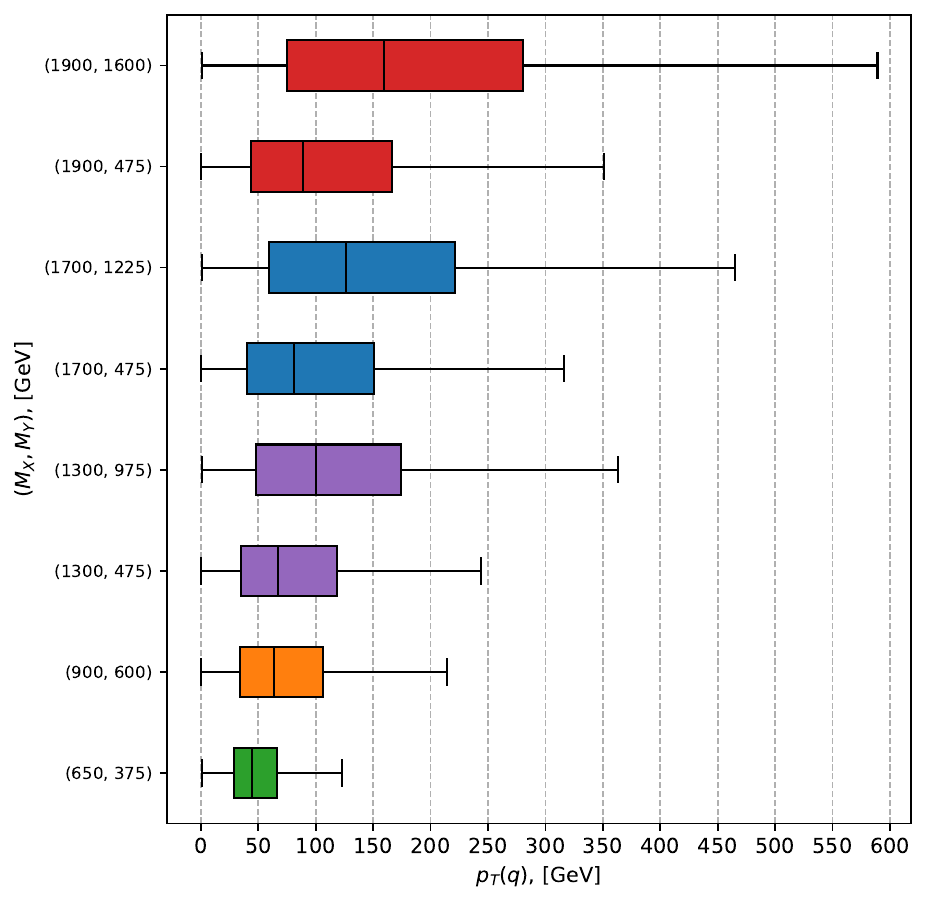}
  \includegraphics[width=0.47\columnwidth]{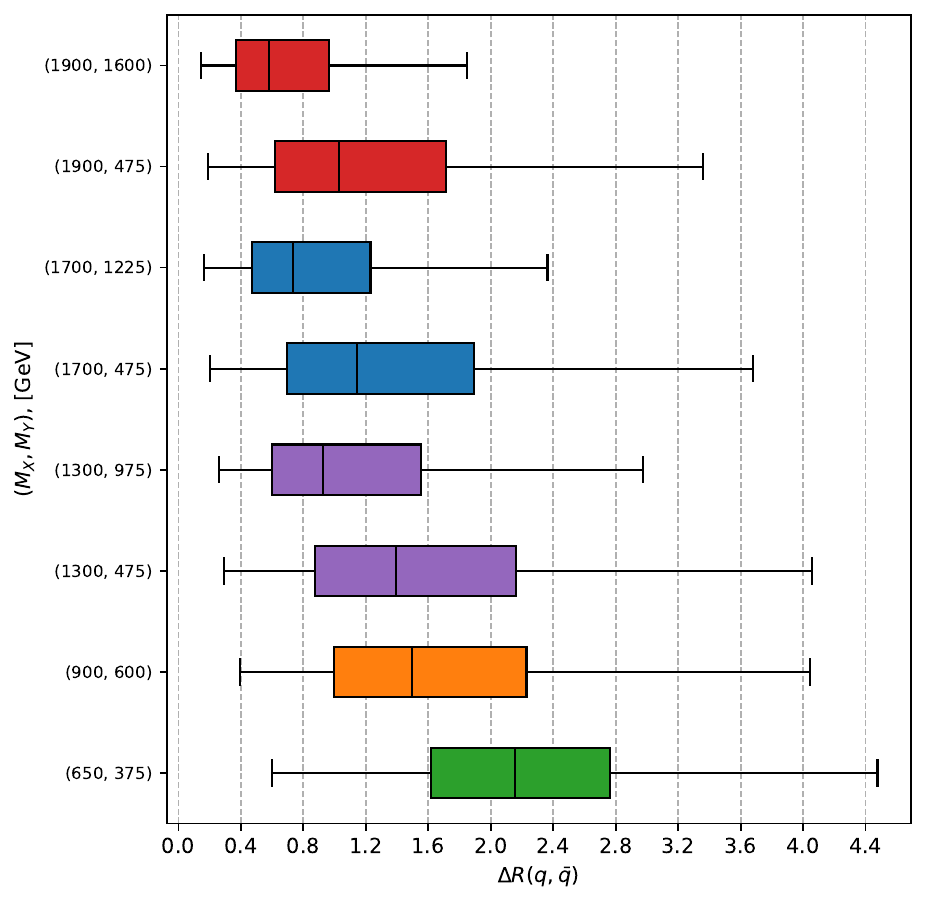} \\
  \caption{ Left: box plots for transverse-momentum distributions of light quarks from $W$ boson decay. Right: $\Delta R = \sqrt{\Delta \varphi^2 + \Delta \eta^2}$ between quarks from $W$ boson decay. Results are given for different mass scenarios of $X$ and $Y$ resonances. }
  \label{fig:fcc_q}
\end{figure}

\begin{figure}
  \centering
  \includegraphics[width=0.50\columnwidth]{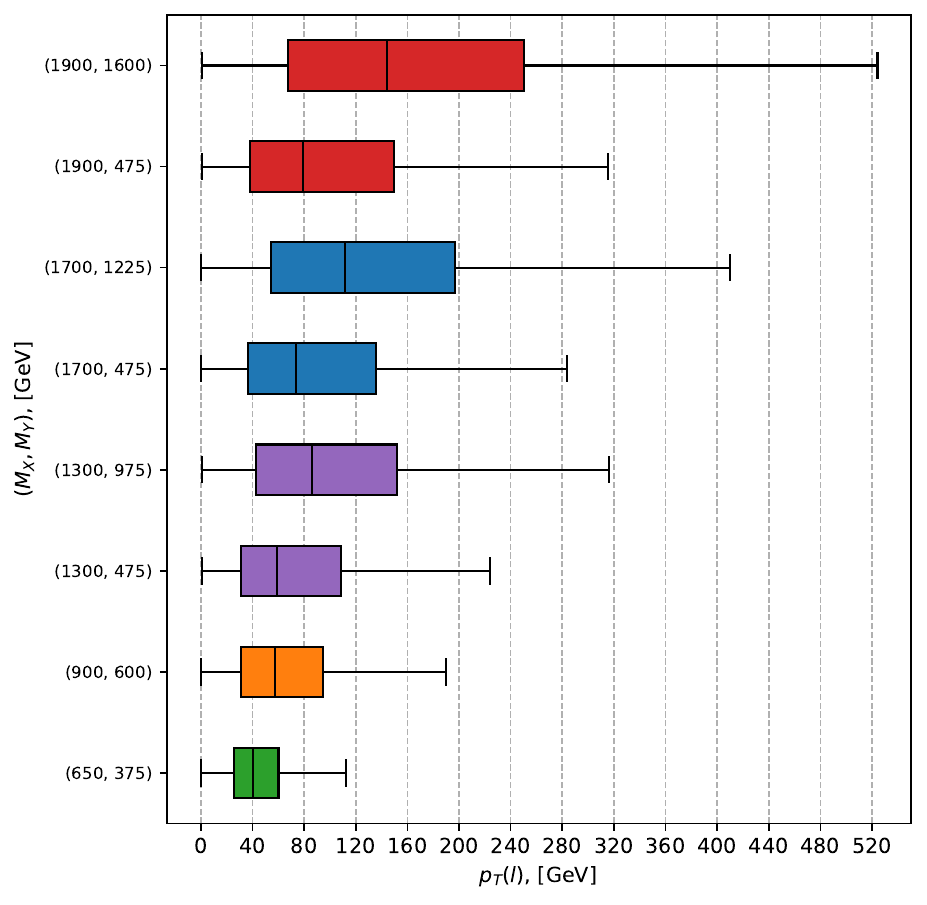}
  \includegraphics[width=0.46\columnwidth]{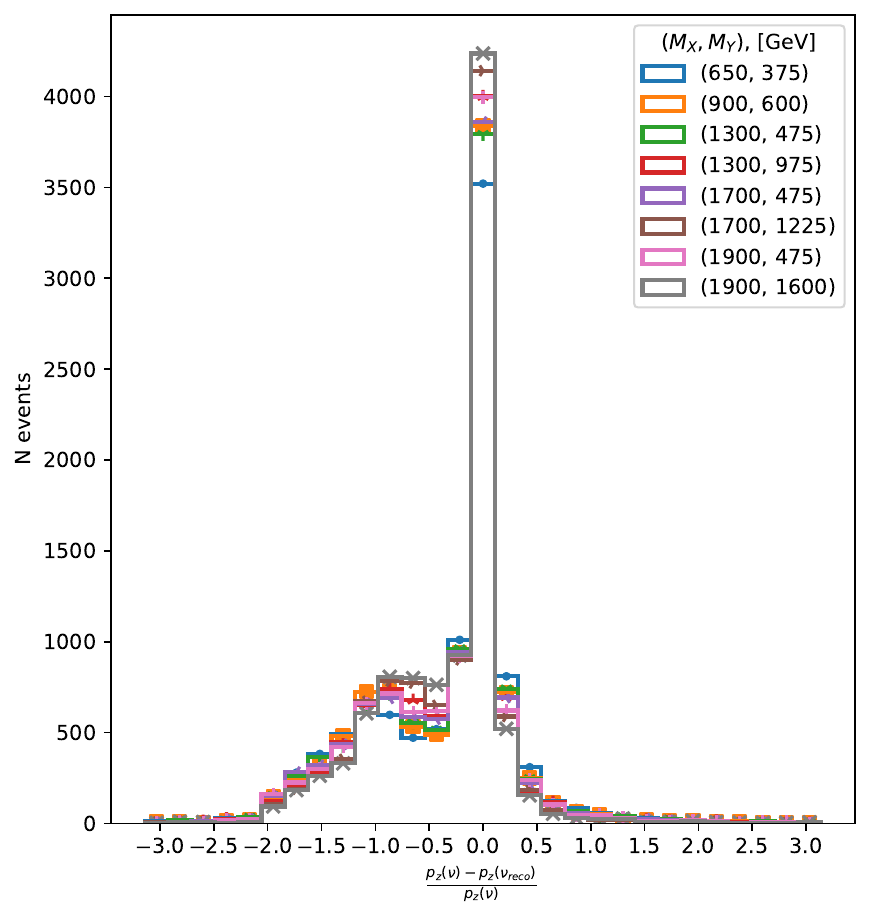} \\
  \caption{ Left: box plots for transverse-momentum distributions of charged lepton ($e$ or $\mu$) from $W$ boson decay. Right: mismatch between generated neutrino longitudinal momentum and reconstructed as described in Sec. \ref{section_b1}. Results are given for different mass scenarios of $X$ and $Y$ resonances. }
  \label{fig:fcc_lnu}
\end{figure}

Box plots for distributions of transverse-momentum of $W$ bosons from top quarks decays are given in Fig. \ref{fig:fcc_bot2} (left). The features with defined regions of events distributions such as $\Delta \varphi$ between $W$ bosons (Fig. \ref{fig:fcc_W}, right) could be used for separation of signal against background events. 
The reconstruction of $W$ bosons with hadronic decay could be done from two quarks tagged as light flavored jets (neglecting the contribution of possible $b$ quarks production from $W$ decay).
This light quarks are also highly energetic populating regions mainly from $30$ GeV to $270$ GeV (Fig. \ref{fig:fcc_q}, left) for considered mass scenarios. 
They are well separated in $\Delta R = \sqrt{\Delta \varphi^2 + \Delta \eta^2}$ (Fig. \ref{fig:fcc_q}, right) and could be reconstructed as two different jets.
The reconstruction of $W$ bosons with leptonic decay can not be performed directly due to the undetected neutrino momentum. Instead, missing transverse momentum $p_T^{MET}$ computed 
as the negative of the vector $p_T$ sum of all reconstructed particles could be taken as approximate value of transverse neutrino momentum $p_T^{\nu}$.
The neutrino longitudinal momentum is computed by solving a quadratic equation in $p_z^{MET}$, employing the four-momenta of the lepton and $W$ boson, $p_T^{\nu}$ and the $m_W = 80$
GeV constraint on the $W$ boson mass. Mismatch between generated and reconstructed neutrino longitudinal momentum based on parton-level information is shown at Fig. \ref{fig:fcc_lnu} (right).

Because of the computing power limitation for the detector-level simulation we consider reduced set of mass points,
selected to have most distinct LHE level kinematic variables distributions.
E.g. from following considered mass points $(m_X, m_Y)$ = (1900, 475), (1900, 600), (1900, 725), (1900, 850), (1900, 975), (1900, 1100), (1900, 1225), (1900, 1350), (1900, 1475), (1900, 1600),
transverse-momentum distributions of b-quarks from top quarks pair decays of (1900, 475) and (1900, 1600) mass points are best separated,
while transverse-momentum distributions of b-quarks for other points located in between. Similar separation behaviour is observed for other considered variables.

\subsection{ Detector-level } \label{section_b2}
In order to accurately reproduce the realistic conditions of the LHC analysis, we apply following object selections (same as in SM $t\bar{t}H$ searches \cite{CMS:2018hnq}):
\begin{itemize}
  \item electron (muon) candidates are required to have $p_T > 30$ GeV ( $p_T > 26$ GeV ) and $| \eta |  < 2.1$; for the selected electron (muon) candidates the tracking efficiency of the CMS detector is varied from 0.83 to 0.95 (0.98 to 0.99).
  \item electron candidates in the transition region between the barrel and endcap calorimeters, $1.4442 < | \eta | < 1.556$, are excluded;
  \item electron (muon) candidates are selected if they have values of relative isolation discriminant $I_{rel} < 0.06$ ($I_{rel} < 0.15$);
  \item jets, reconstructed by anti-$k_T$ algorithm with a distance parameter of $0.4$, are required to have $p_T > 30$ GeV and $| \eta |  < 2.4$;
\end{itemize}
and following events selections:
\begin{itemize}
  \item events are required to have at least two b-tagged jets (at ``loose''\footnotemark working point to match Delphes parameterization \cite{CMS:2012feb}, defined by 10\% rate for misidentifying a light jet as a b jet);
  \item events are required to fulfill $p_T^{MET} > 20$ GeV condition;
  \item events with additional isolated selected leptons with $p_T > 15$ GeV are excluded from further analysis.
\end{itemize}
After application of the selections the main backgrounds are known to be $t\bar{t}$ (93\% of total background events), single top quark production (4\%), $W/Z + jets$ (2\%), $t\bar{t}+Z/W$ (0.4\%), SM $t\bar{t}+H$ (0.2\%) and diboson production process (0.01\%) \cite{CMS:2018hnq}. Background contributions from QCD multijet production is negligible.
Thus, for the following study we are considering $t\bar{t}$ production process as main background of interest.

\begin{figure}
  \centering
  \includegraphics[width=0.47\columnwidth]{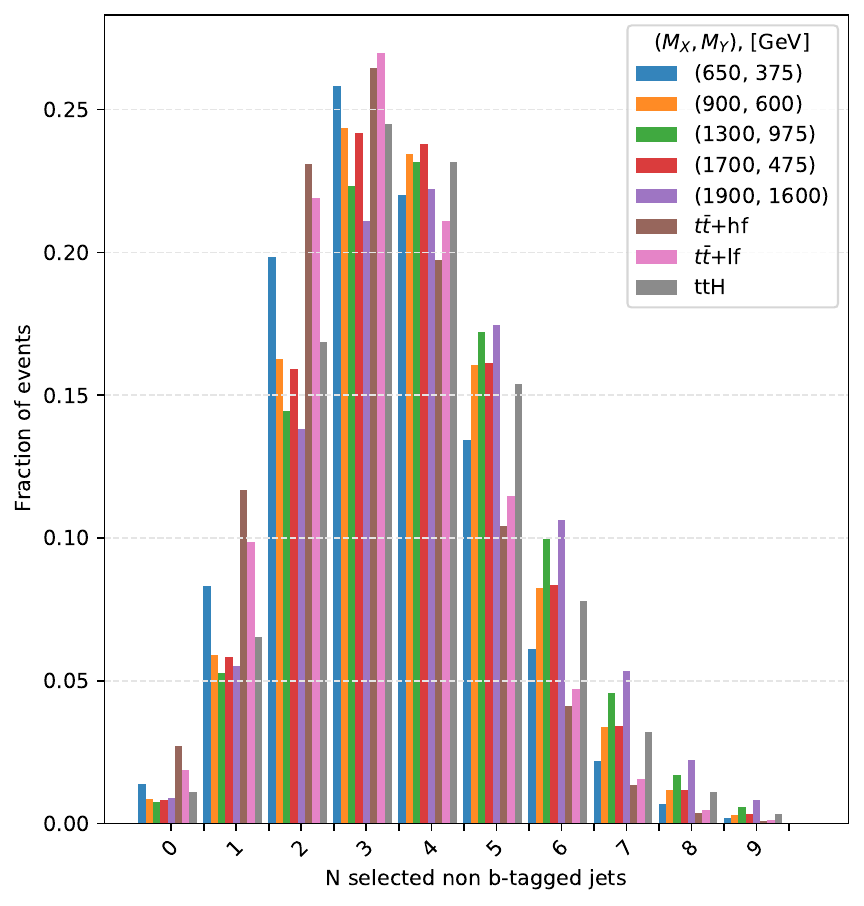}
  \includegraphics[width=0.47\columnwidth]{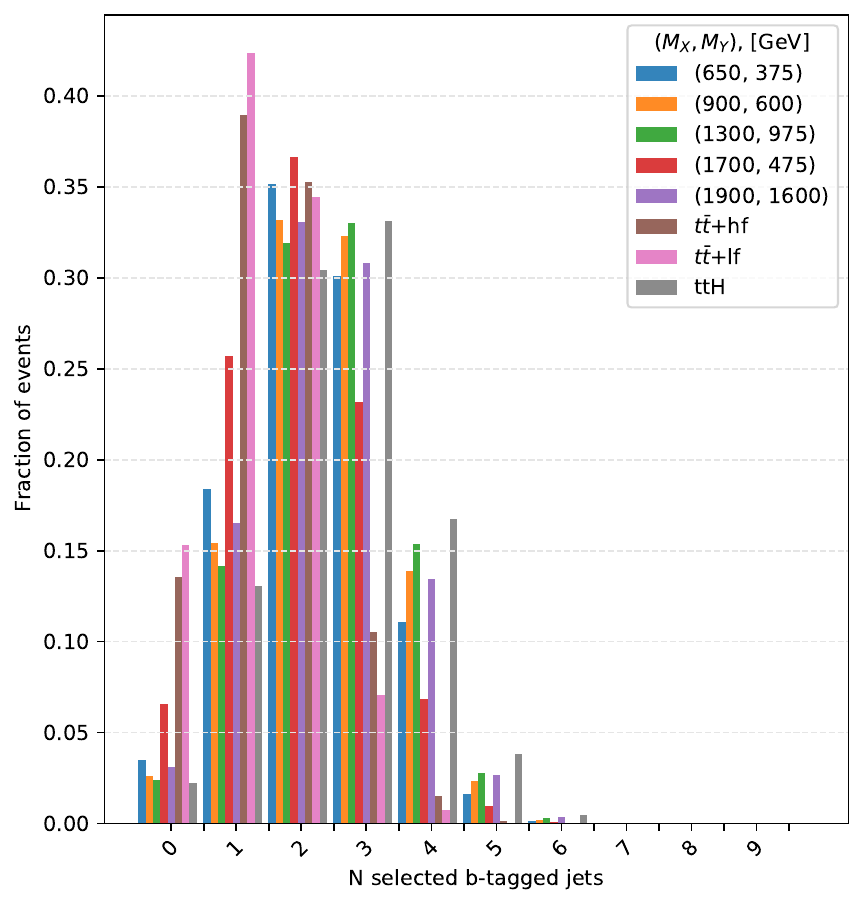} \\
  \includegraphics[width=0.47\columnwidth]{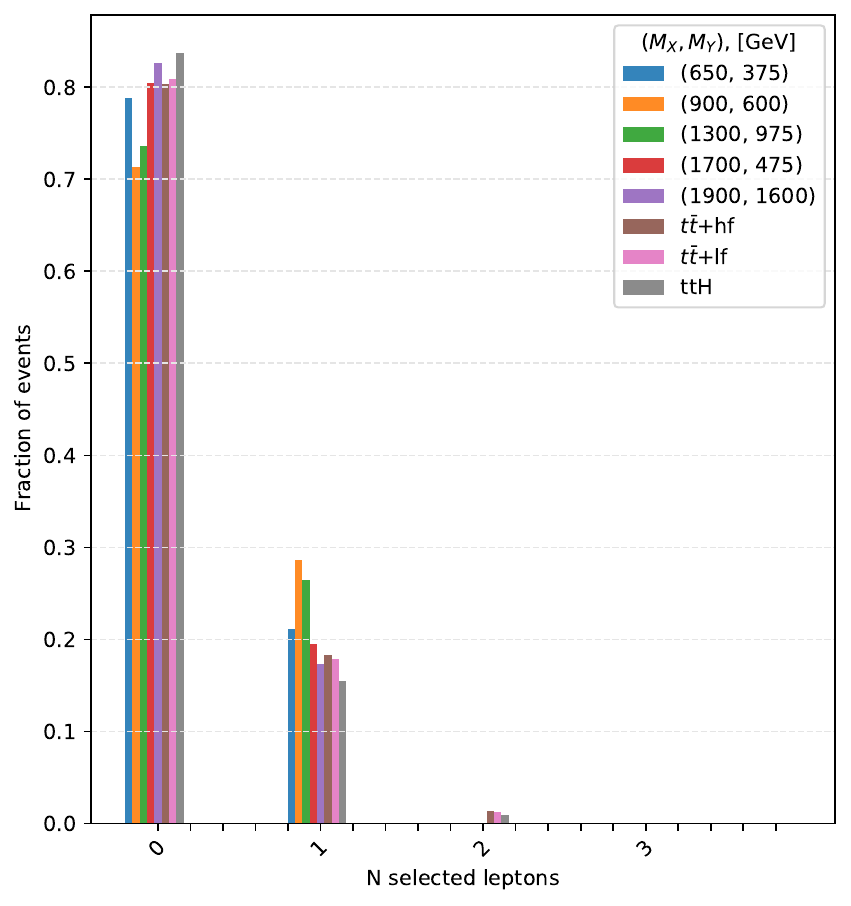}
  \includegraphics[width=0.42\columnwidth]{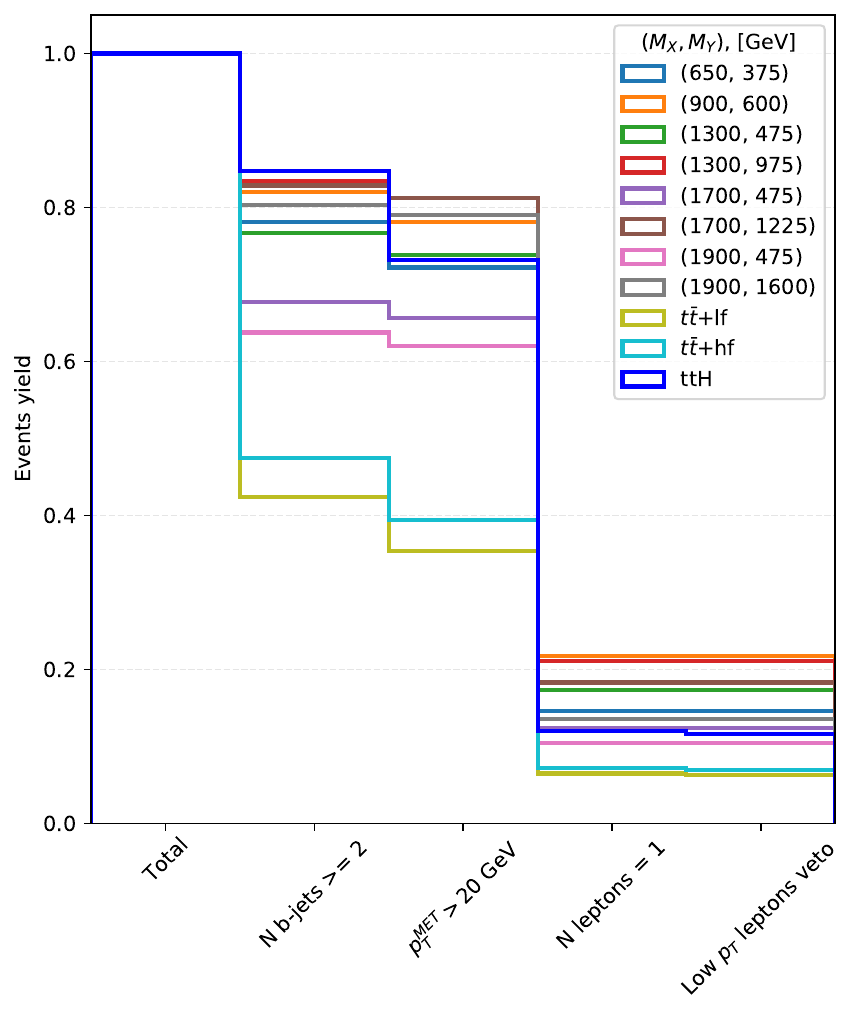} \\
  \caption{ Number of selected objects: light jets (top-left), b-tagged jets (top-right), leptons (bottom-left), and fraction of events after baseline selections (bottom-right). Results are given for signal process for different mass scenarios of $X$ and $Y$ resonances and for SM $t\bar{t}$ background. }
  \label{fig:mass_points}
\end{figure}

Signal samples show comparable distributions for different mass points (Fig. \ref{fig:mass_points}).
The number of jets is enhanced in energetic events with resonant signal production in comparison to SM $t\bar{t}$ process.
The number of b-tagged jets is increased with increasing mass of $Y$ and $X$ resonances, but limited by boosted regime for heavy masses of  $X$ and light masses of $Y$.
The signal events selection rates are between of $20\%$ for $m_X,m_Y = (1300, 975)$ GeV mass point and $10\%$ for $(1900, 475)$ GeV mass point (Fig. \ref{fig:mass_points} and Tab. \ref{tab:sel_eff}).
Resonant production nature of $t\bar{t}$ pair in signal samples consequently leads to enhanced $p_T$ distributions of leptons (comparing to SM $t\bar{t}H$ sample) and increasing selection efficiency for heavy mass points. On the other hand, isolated lepton selection efficiencies vary over mass points due to the difference in separation
of charged lepton and hadronic decay products. 
For heavy $Y$ resonance boosted top quark decay is enhanced, while for heavy $X$ and low mass $Y$ the decay of the latter is boosted and top-quarks products are less separated. 
Thus, highest efficiency observed for medium mass points.
We found selection efficiency for $t\bar{t}$+lf and $t\bar{t}$+hf to be in agreement with SM $t{t}H$ searches \cite{CMS:2018hnq} efficiency when ``medium'' b-tagging working point is applied 
and greater for `loose'' working point we use.

\footnotetext{``loose'', ``medium'' and ``tight'' working point values of b-tagging discriminator threshold are defined by misidentification probability for light-parton jets close to 10\%, 1\% and 0.1\% respectively, at an average jet $p_T$ of about 80 GeV/c \cite{CMS:2012feb}.}

\begin{table}
\begin{center}
\begin{scriptsize}
\begin{tabular}{c|c|c|c|c}
               & \multicolumn{4}{c}{Selection} \\
    Process    &  $N_{\text{b-jets}} >= 2$ &  $p_T^{MET} > 20$ GeV &  $N_{\text{leptons}} = 1$ &  Low $p_{T}$ leptons veto \\ \hline
    (650, 375) &          0.781 &                   0.722 &          0.147 &                     0.147 \\
    (900, 600) &          0.820 &                   0.781 &          0.217 &                     0.217 \\
   (1300, 475) &          0.767 &                   0.739 &          0.173 &                     0.173 \\
   (1300, 975) &          0.834 &                   0.812 &          0.211 &                     0.211 \\
   (1700, 475) &          0.677 &                   0.656 &          0.124 &                     0.124 \\
  (1700, 1225) &          0.828 &                   0.812 &          0.183 &                     0.183 \\
   (1900, 475) &          0.638 &                   0.620 &          0.105 &                     0.105 \\
  (1900, 1600) &          0.804 &                   0.790 &          0.136 &                     0.136 \\ \hline
   $t\bar{t}H$ &          0.847 &                   0.732 &          0.120 &                     0.117 \\
 $t\bar{t}$+lf &          0.424 &                   0.354 &          0.065 &                     0.063 \\
 $t\bar{t}$+hf &          0.475 &                   0.395 &          0.072 &                     0.069 \\
\end{tabular}
\end{scriptsize}
  \caption{ \label{tab:sel_eff} fraction of events after sequential application (from left to right) of baseline selections for signal mass $(M_X, M_Y)$ benchmark points and backgrounds. }
\end{center}
\end{table}

\begin{table}
\begin{center}
\begin{scriptsize}
\begin{tabular}{c|c|c|c|c}
               & \multicolumn{4}{c}{Selection} \\
    Process    &  $N_{\text{b-jets}} >= 2$ &  $p_T^{MET} > 20$ GeV &  $N_{\text{leptons}} = 1$ &  Low $p_{T}$ leptons veto \\ \hline
 $t\bar{t}$+lf &       36874808 &                30794313 &        5645613 &                   5467264 \\
 $t\bar{t}$+hf &       14991361 &                11746426 &        1395587 &                   1374814 \\
           ttH &          58834 &                   50873 &           8328 &                      8107 \\
\end{tabular}
\end{scriptsize}
  \caption{ \label{tab:sel_eff2} expected number of events after sequential application (from left to right) of baseline selections for 
             luminosity of 137 fb$^{-1}$ of proton-proton collisions collected by the CMS detector during Run II. }
\end{center}
\end{table}

\begin{figure}
  \centering
  \includegraphics[width=0.47\columnwidth]{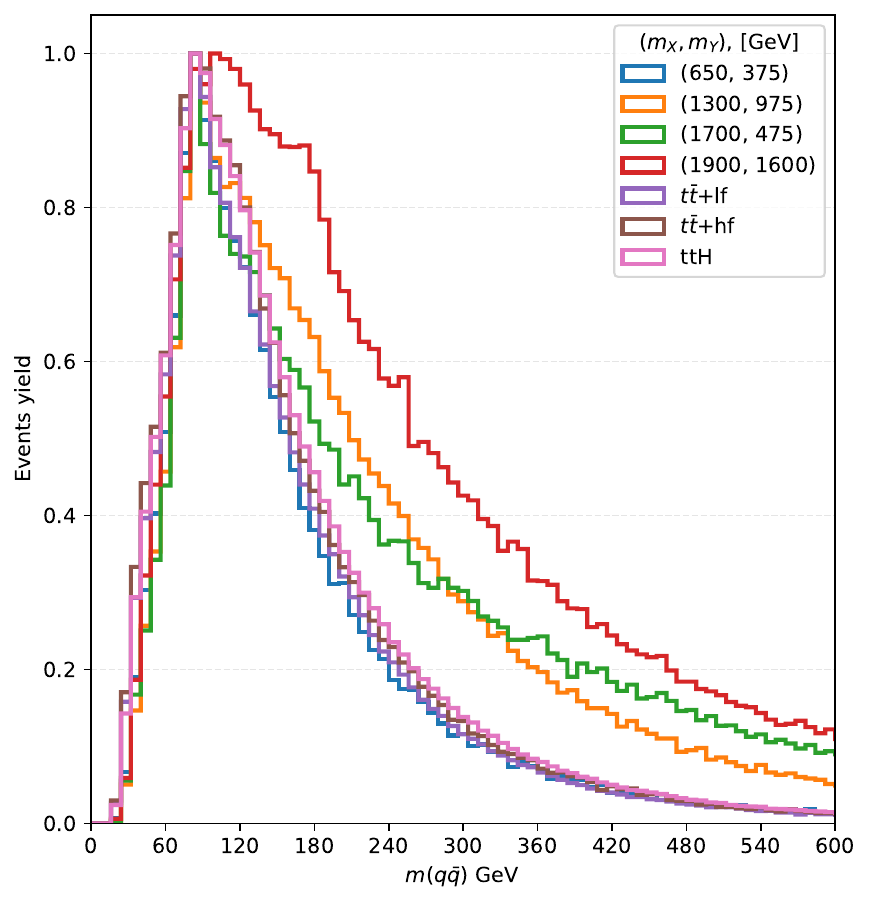}
  \includegraphics[width=0.47\columnwidth]{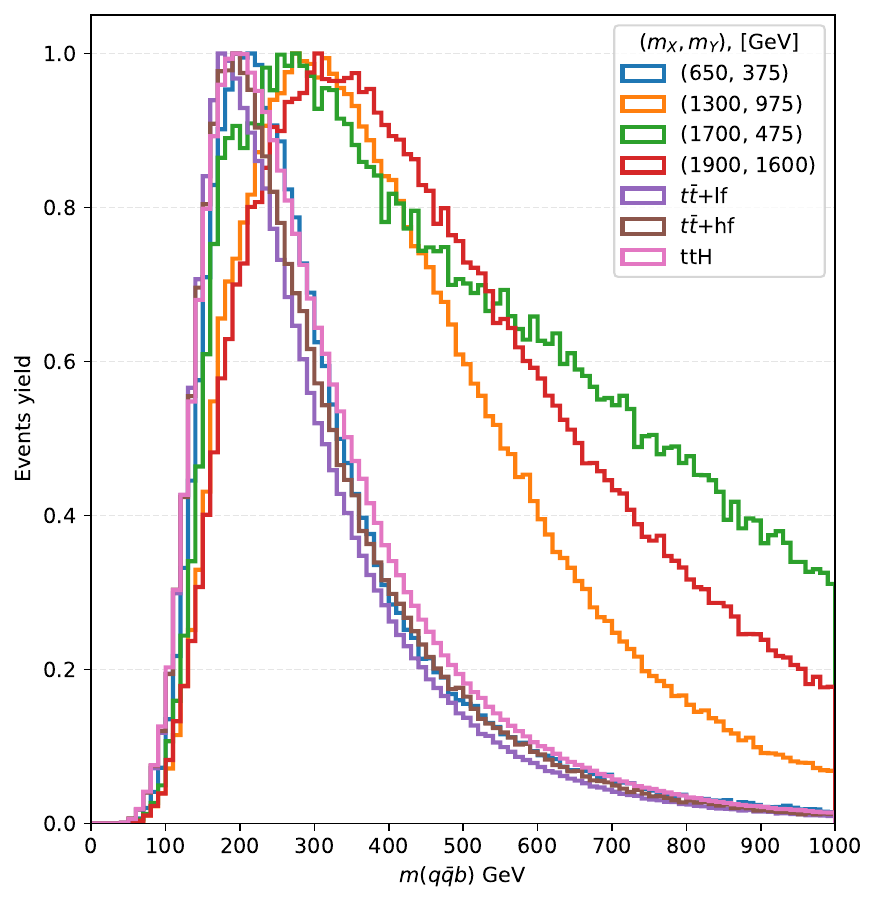} \\
  \includegraphics[width=0.31\columnwidth]{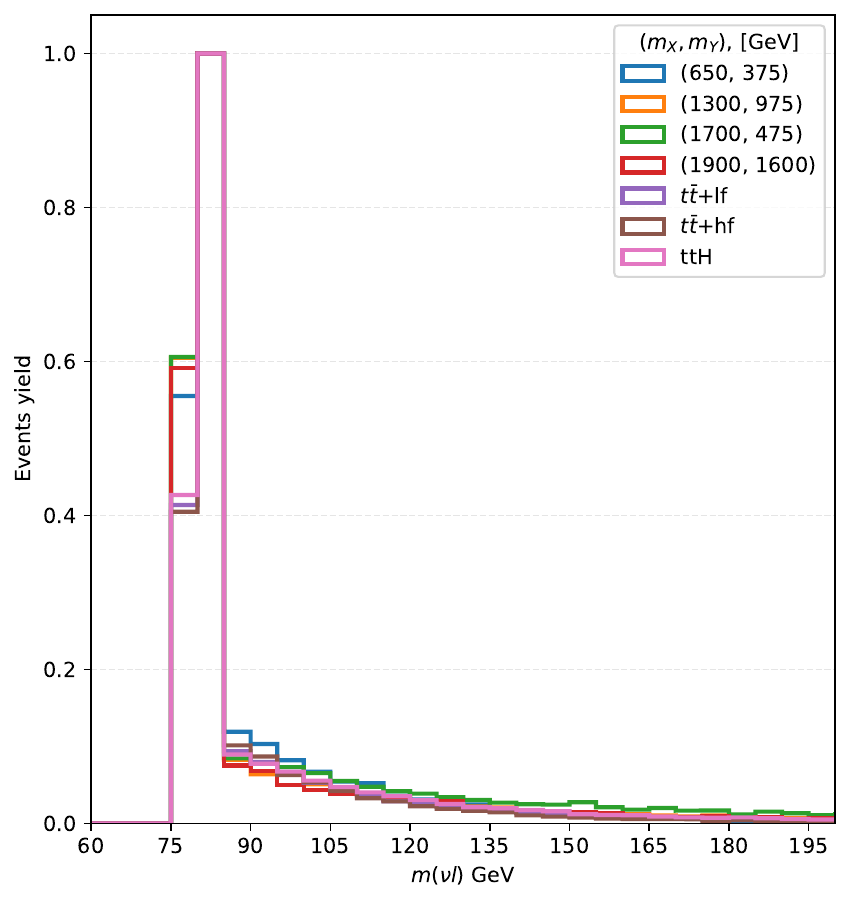}
  \includegraphics[width=0.31\columnwidth]{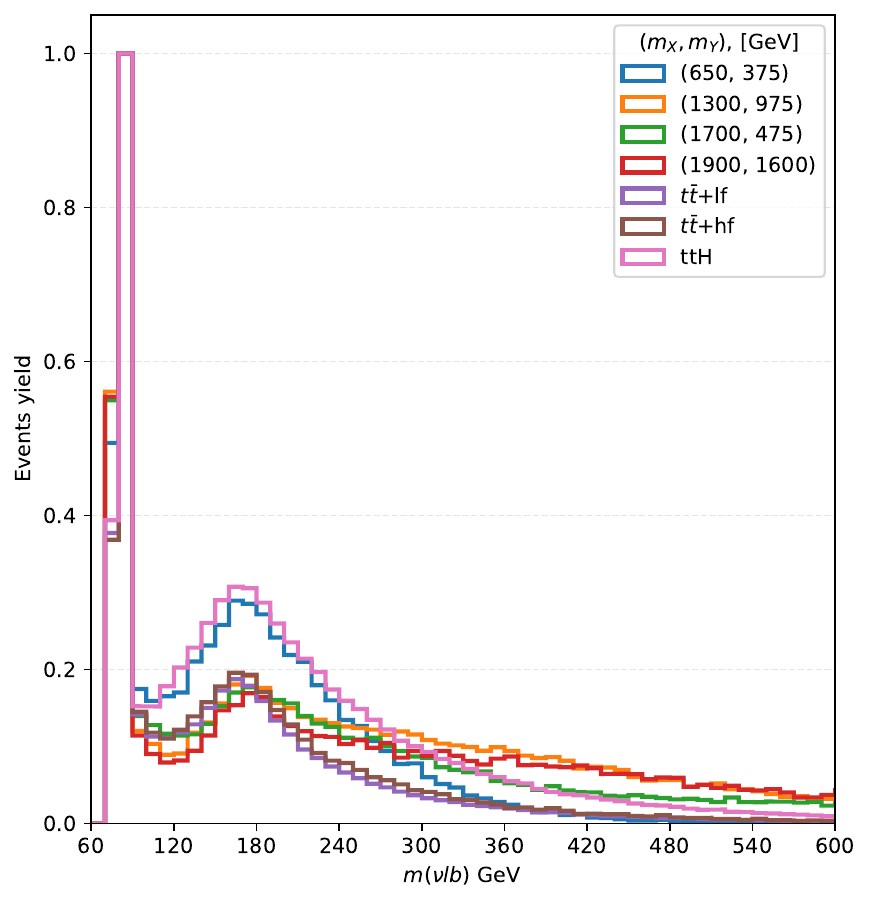}
  \includegraphics[width=0.31\columnwidth]{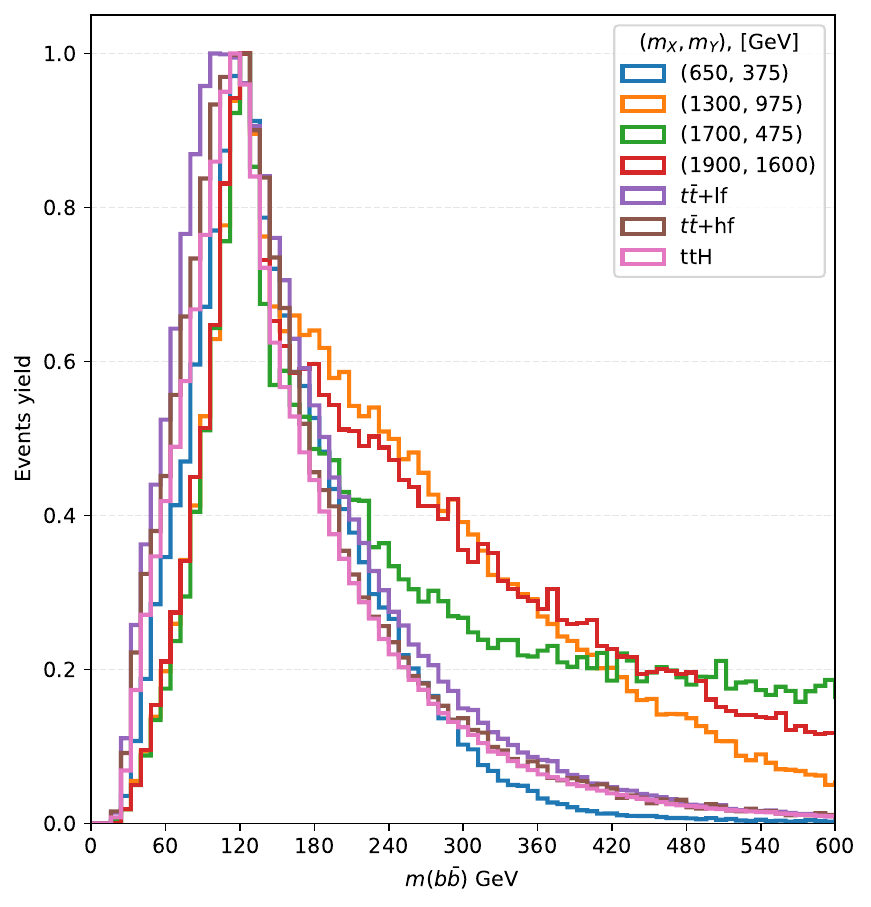} \\
  \caption{ Invariant masses reconstructed from all possible combinations of reconstructed and selected objects: pair of light jets (hadronic $W$ boson candidate, top-left), pair of light jets and b-tagged jet (hadronic top-quark candidate, top-right), neutrino and charged lepton (leptonic $W$ boson candidate, bottom-left), neutrino and charged lepton and b-tagged jet (leptonic top-quark candidate, bottom-center), pair of b-tagged jets (Higgs candidate, bottom-right). Results are given for signal process for different mass scenarios of $X$ and $Y$ resonances and for SM $t\bar{t}$ background. }
  \label{fig:mass_combo_0}
\end{figure}

\begin{figure}
  \centering
  \includegraphics[width=0.47\columnwidth]{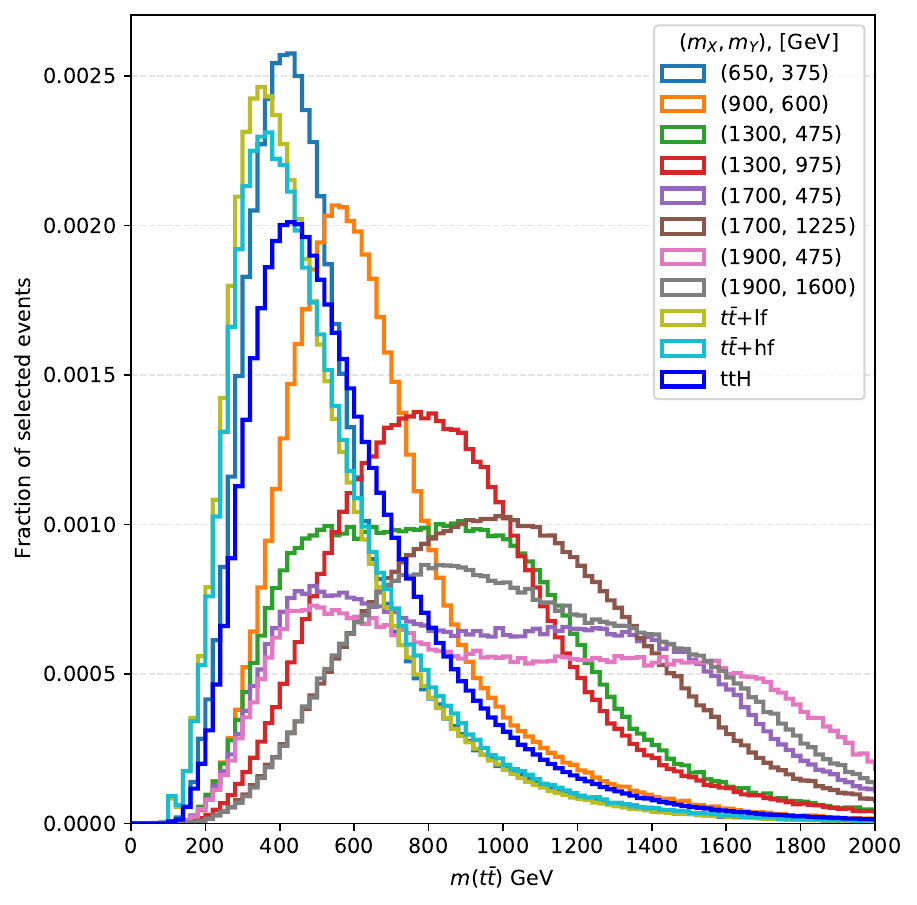}
  \includegraphics[width=0.47\columnwidth]{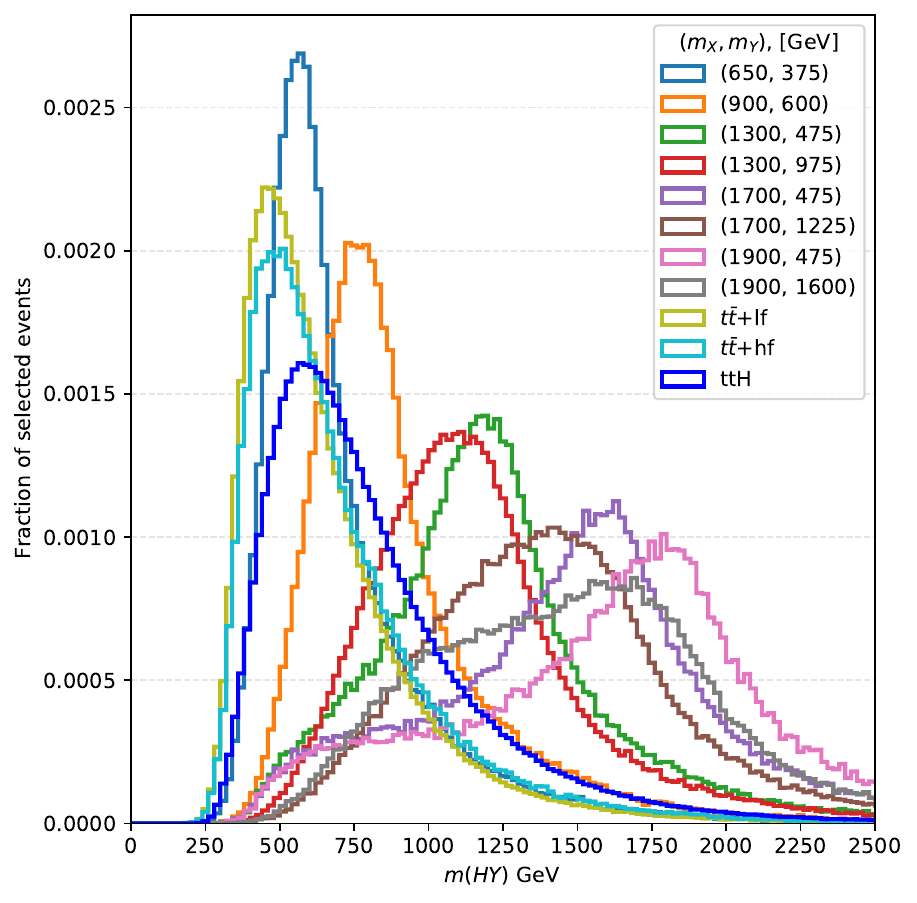} \\
  \caption{ Invariant masses reconstructed from all possible possible combinations of reconstructed objects: pair of top-quarks candidates ($Y$ candidate, left) and Higgs and $Y$ candidates ($X$ candidate, right). Results are given for signal process for different mass scenarios of $X$ and $Y$ resonances and for SM $t\bar{t}$ background. }
  \label{fig:mass_combo}
\end{figure}

To investigate the possibility of $Y$ and $X$ reconstruction we process over selected objects of event to create all possible unique sets with following content:
\begin{eqnarray}
 C_{\textrm{full}} = \{ b_{t}^{l}, ~l, ~\nu,~ b_{t}^{q}, ~q^{1}_{t}, ~q^{2}_{t},~ b^{1}_{H}, ~ b^{2}_{H} \}
\label{rat3}
\end{eqnarray}
where $b_{t}^{l}$ is supposed to be b-tagged jet from leptonic top-quark decay, $b_{t}^{q}$ - b-tagged jet from hadronic top-quark decay, $q^{1}_{t}$ and $q^{2}_{t}$ - $p_T$ leading and subleading light jets (non-b-tagged) from  hadronic top-quark decay, $b^{1}_{H}$ and $b^{2}_{H}$ - $p_T$ leading and subleading b-tagged jets from $H$ decay.
The obtained sets of objects are used to reconstruct the kinematic of events and fill invariant masses histograms shown at Fig. \ref{fig:mass_combo_0} and Fig. \ref{fig:mass_combo}.
The peaks from W-bosons and top-quarks decays are visible in signal $q\bar{q}$, $\ell\nu$, $q\bar{q}n$ and $\ell\nu b$ distributions. However, they could not provide a well separation from $t\bar{t}$ SM background.
The distributions for heavy resonant masses are also characterized by resolution degradation.
On the other hand, the signal is clearly distinguishable in $t\bar{t}$ and $HY$ candidates distributions with histograms peaks near the values of the corresponding masses of resonances.

\begin{figure}
  \centering
  \includegraphics[width=0.47\columnwidth]{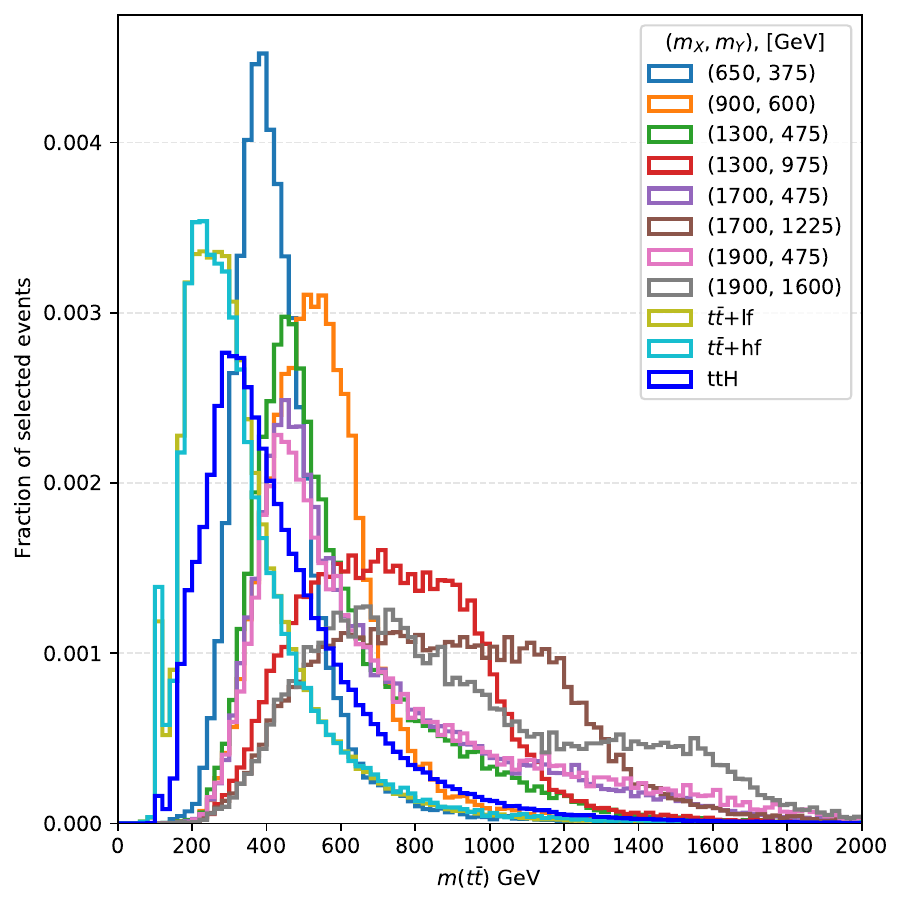}
  \includegraphics[width=0.47\columnwidth]{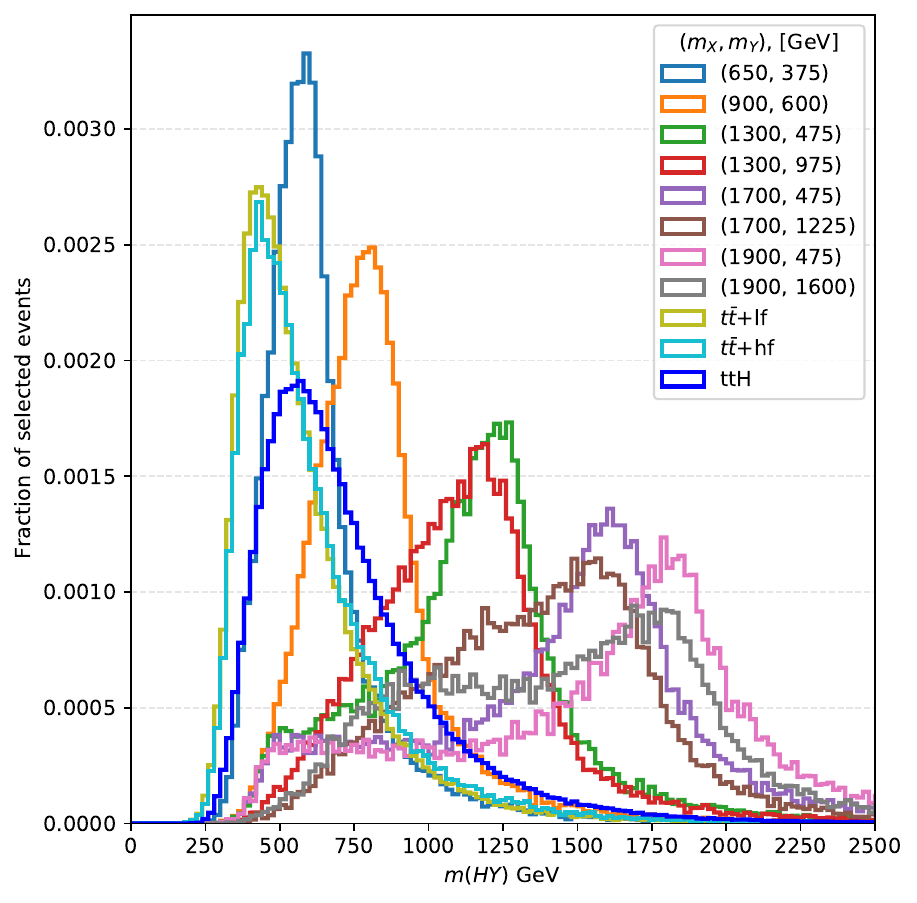} \\
  \caption{ Invariant masses reconstructed from candidates with lowest metric eq. (\ref{metric}) value: pair of top-quarks candidates ($Y$ candidate, left) and Higgs and $Y$ candidates ($X$ candidate, right). Results are given for signal process for different mass scenarios of $X$ and $Y$ resonances and for SM $t\bar{t}$ background. }
  \label{fig:mass_combo3}
\end{figure}

Now, when the possibility to reconstruct signal signature is shown, the target of the analysis is to define optimal (e.g. from the point of view of kinematic variables resolution) 
selection rule to choose only one set (\ref{rat3}) in the event. 
The most common technique to score a permutation set is a $\chi^2$-minimization based on the consistency of the reconstructed masses with known values.
We probe a universal over all mass points and independent from $X$ and $Y$ resonances masses $\chi^2$ metric:
\begin{eqnarray}
 \chi^2 = \Big(\frac{ m_{qq} - m^{SM}_{W} }{  \sigma(m_{qq}) }\Big)^2 + \Big(\frac{ m_{\ell\nu} - m^{SM}_{W} }{  \sigma(m_{\ell\nu}) }\Big)^2 +  \nonumber  \\
   + \Big(\frac{ m_{bqq} - m^{SM}_{t} }{ \sigma(m_{bqq}) }\Big)^2 + \Big(\frac{ m_{b\ell\nu} - m^{SM}_{t} }{  \sigma(m_{b\ell\nu}) }\Big)^2
   + \Big(\frac{ m_{bb} - m^{SM}_{H} }{ \sigma(m_{bb}) }\Big)^2 
\label{metric}
\end{eqnarray}
where $m^{SM}_{W}$, $m^{SM}_{t}$ and $m^{SM}_{H}$ are the SM values of the masses and $\sigma(..)$ are mass resolutions extracted from respective distributions.
Invariant masses histograms filled by objects from sets (\ref{rat3}) with lowest metric (\ref{metric}) value in the event are available at Fig.~\ref{fig:mass_combo3} showing moderate improvement in mass resolution.

\subsection{ Deep Neural Network application } \label{section_b3}
The problem to define optimal selection rule to choose only one set (\ref{rat3}) per event could be considered by using Machine Learning techniques.
Unlike the regression models widely used in experimental high energy physics (HEP) to separate signal and background events, 
in our case we need to assign reconstructed jet to the top-quarks or Higgs boson decay chains.
We considered several options to apply Machine Learning approach to this problem.
First of all, the definition of the jets assignment problem is similar to reconstruction (clustering) tasks \cite{Shlomi:2020ufi, 8614089, Ju:2020xty, Thais:2022iok, Fenton:2020woz} 
where Graph Neural Networks (GNN) inspired architectures were applied.
The target of GNN is to establish edges or connections between input points.
When individual connections are irrelevant
the output of the such network could be considered as a set of values per constituent representing the credibility of the constituent to be a part of a cluster (set).

In our implementation the neural network input is a list of selected jets, represented by their 4-vector (as transverse momentum, pseudorapidity, azimuthal angle and mass) and boolean b-tagging value.
In additional, we add following event features providing extra information about reconstructed kinematic: invariant mass of every jet's pair,
neutrino transverse momentum and azimuthal angle, transverse momentum, pseudorapidity and azimuthal angle and kind of selected lepton 
and 4-vector of $W$ boson reconstructed from leptonic decay.
In order to increase the available statistics, the training set is prepared without events selections from Section \ref{section_b2}.
The number of jets is limited to $N_{jets}^{DNN} = 8$ ordering in $p_T$. For events with number of jets $< N_{jets}^{DNN}$ the extra features are filled with 0, 
as well as features of lepton and $W$ boson in the events without selected lepton.
The output of the network is a set of values per jet representing the credibility of the jet to be a part of $H$ boson, leptonic top-quark or hadronic top-quark cluster.
The desired output value of training sample with a correct assignments of the jets is defined using parton generator level information.
Jets are matched to the simulated truth quarks ($b$-quarks from $H$ boson and top-quark, light quarks from $W$ boson decays) using $\Delta R = \sqrt{\Delta \eta^2 + \Delta \phi^2} < 0.4.$ criterion.
{\scshape TensorFlow} and {\scshape Keras} \cite{tensorflow2015-whitepaper, chollet2015keras} packages are used for the definition, training and evaluation of the Deep Neural Network model, 
separately for each mass point.
Sequential DNN model with five hiden layers is chosen with parameters selected by trial and error. 
The number of hidden neurons started at twice the size of the input layer and decreased to doubled size of the output layer. 
For the hiden layers ``ReLU'' activation function is used while ``Linear'' and ``Sigmoid'' activation functions are used for input and output layers respectively. 
As loss function binary cross entropy is chosen. 
Input features are normalized to a range $[0, 1]$, while the log transformation is also applied to long tail distributions such as $p_T$ of the jets. 
To estimate importance of the input variables shown in the Table \ref{tab:dnn_vars} we use Permutation feature importance measure \cite{https://doi.org/10.48550/arxiv.1801.01489},
calculating the changes in DNN prediction error in the considered sample when tested values of the probed feature are shuffled over the dataset.

\begin{table}
\begin{center}
\begin{scriptsize}
\setlength\tabcolsep{1.0pt}
\begin{tabular}{c|c|c|c|c|c|c|c|c}
DNN input &  (650, 375) &  (900, 600) &  (1300, 475) &  (1300, 975) &  (1700, 475) &  (1700, 1225) &  (1900, 475) &  (1900, 1600) \\
\hline

$\overline{p_T(j)}$                 &         1.3 &         0.9 &          1.0 &          1.0 &          1.0 &           0.9 &          1.0 &           1.0 \\
$\overline{\eta(j)}$                &         0.3 &         0.3 &          0.3 &          0.3 &          0.4 &           0.4 &          0.3 &           0.3 \\
$\overline{\varphi(j)}$                &         0.4 &         0.4 &          0.3 &          0.4 &          0.3 &           0.6 &          0.3 &           0.5 \\
$\overline{m(j)}$                   &         0.7 &         0.5 &          0.7 &          0.5 &          0.7 &           0.6 &          0.8 &           0.6 \\
$\overline{ \textrm{b-tag score} }$ &         2.3 &         2.5 &          1.3 &          1.7 &          1.3 &           1.7 &          1.5 &           1.6 \\
$\overline{ m(j_i,j_j) }$           &         0.9 &         0.5 &          1.0 &          0.7 &          0.9 &           0.7 &          0.9 &           0.7 \\
$\overline{ \Delta R(j_i,j_j) }$    &         0.5 &         1.0 &          0.7 &          1.0 &          0.7 &           1.0 &          0.8 &           1.0 \\
$p_T(\nu)$                          &         0.7 &         0.7 &          0.7 &          1.0 &          0.7 &           0.8 &          1.0 &           1.0 \\
$\varphi(\nu)$                         &         0.4 &         0.6 &          0.3 &          0.8 &          0.4 &           1.6 &          0.4 &           1.5 \\
$p_T(\ell)$                         &         1.8 &         0.6 &          2.5 &          1.8 &          2.3 &           1.7 &          2.6 &           2.0 \\
$\eta(\ell)$                        &         0.2 &         0.2 &          0.4 &          0.3 &          0.6 &           0.5 &          0.2 &           0.2 \\
$\varphi(\ell)$                        &         0.2 &         0.3 &          0.3 &          0.3 &          0.2 &           0.3 &          0.2 &           0.2 \\
$N_{\mu}$                           &         0.9 &         0.3 &          0.9 &          0.8 &          0.6 &           0.5 &          0.3 &           0.6 \\
$N_{e}$                             &         0.7 &         0.6 &          0.8 &          0.6 &          0.5 &           0.4 &          0.6 &           0.3 \\
$p_T(W_{\ell})$                     &         1.8 &         1.4 &          1.4 &          0.8 &          1.3 &           0.6 &          1.3 &           0.6 \\
$\eta(W_{\ell})$                    &         0.4 &         0.3 &          0.3 &          0.4 &          0.3 &           0.4 &          0.2 &           0.4 \\
$\varphi(W_{\ell})$                    &         0.3 &         0.3 &          0.2 &          0.2 &          0.2 &           0.2 &          0.2 &           0.2 \\
$m(W_{\ell})$                       &         2.2 &         0.9 &          2.5 &          1.2 &          2.9 &           0.8 &          3.1 &           0.8 \\

\end{tabular}
\end{scriptsize}
  \caption{ \label{tab:dnn_vars} importance score of the DNN classifier input features. Average values are given for jet related variables. Scores are normalized within mass points. }
\end{center}
\end{table}

\begin{figure}
  \centering
  \includegraphics[width=0.47\columnwidth]{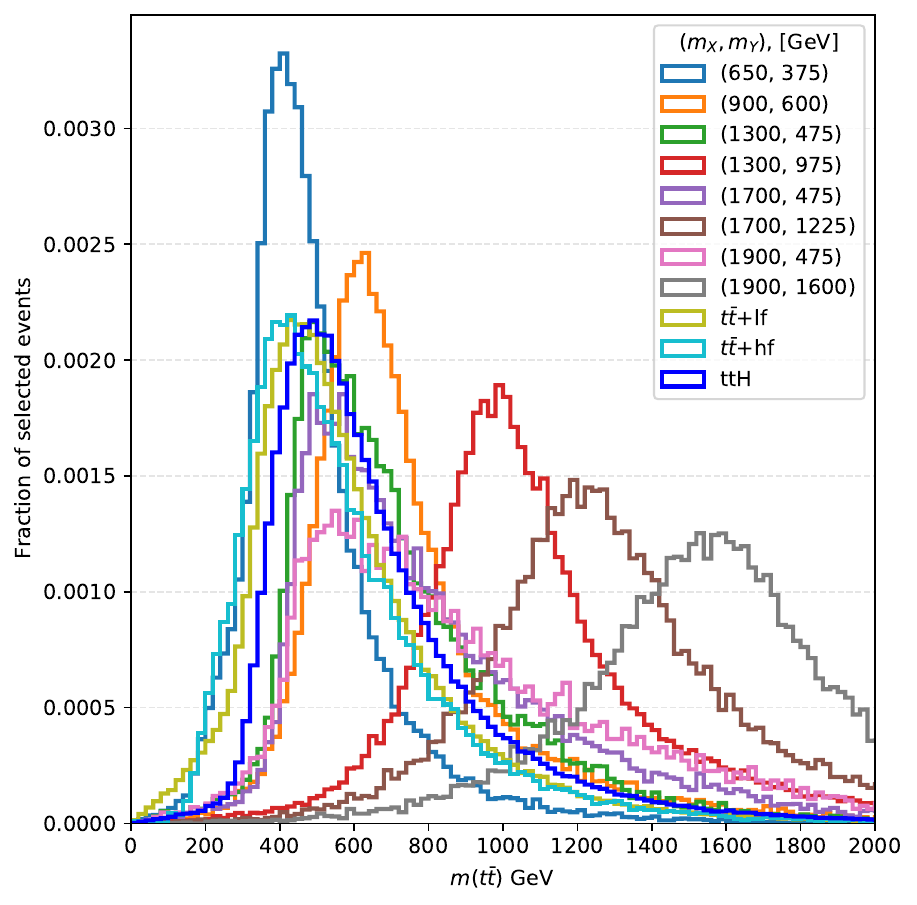}
  \includegraphics[width=0.47\columnwidth]{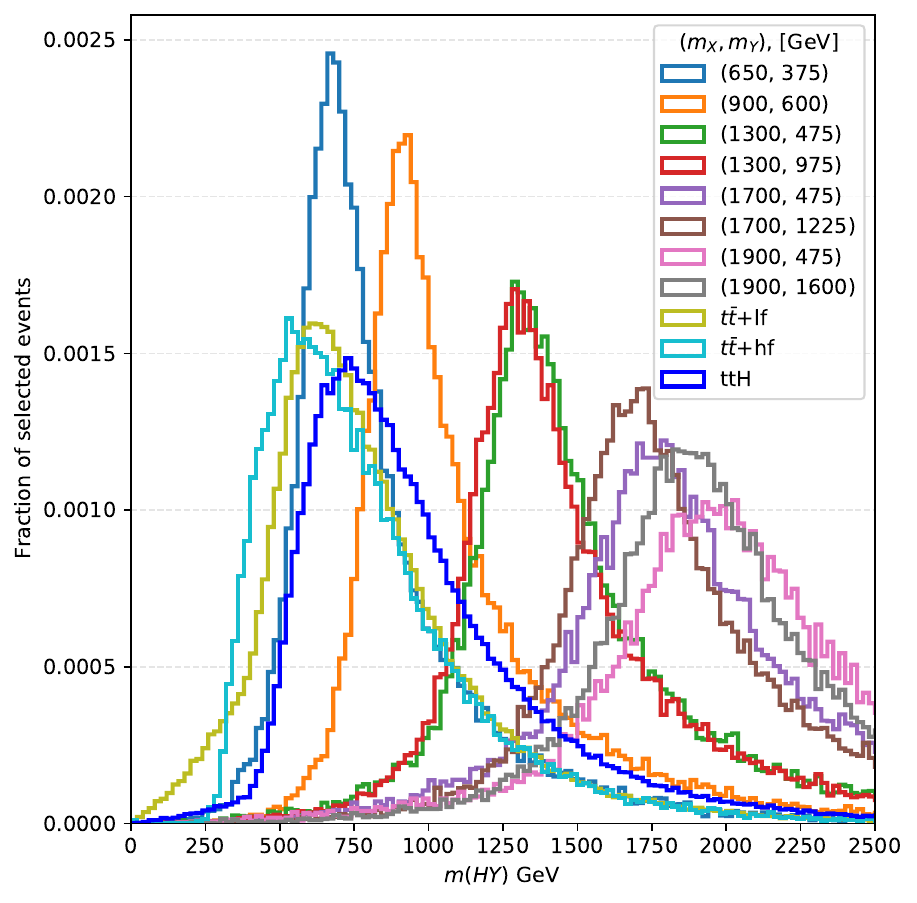} \\
  \caption{ Invariant masses reconstructed from jets selected using DNN score: pair of top-quarks candidates ($Y$ candidate, left) and Higgs and $Y$ candidates ($X$ candidate, right). Results are given for signal process for different mass scenarios of $X$ and $Y$ resonances and for SM $t\bar{t}$ background. }
  \label{fig:mass_combo_eval}
\end{figure}

\begin{figure}
  \centering
  \includegraphics[width=0.47\columnwidth]{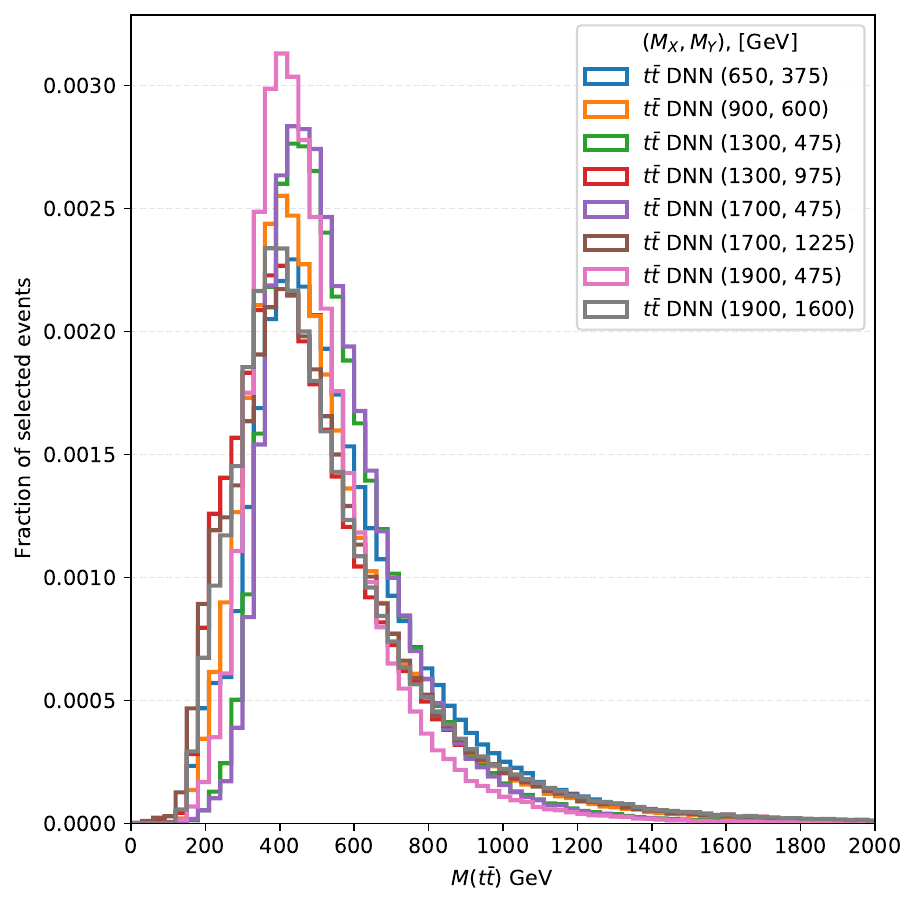}
  \includegraphics[width=0.47\columnwidth]{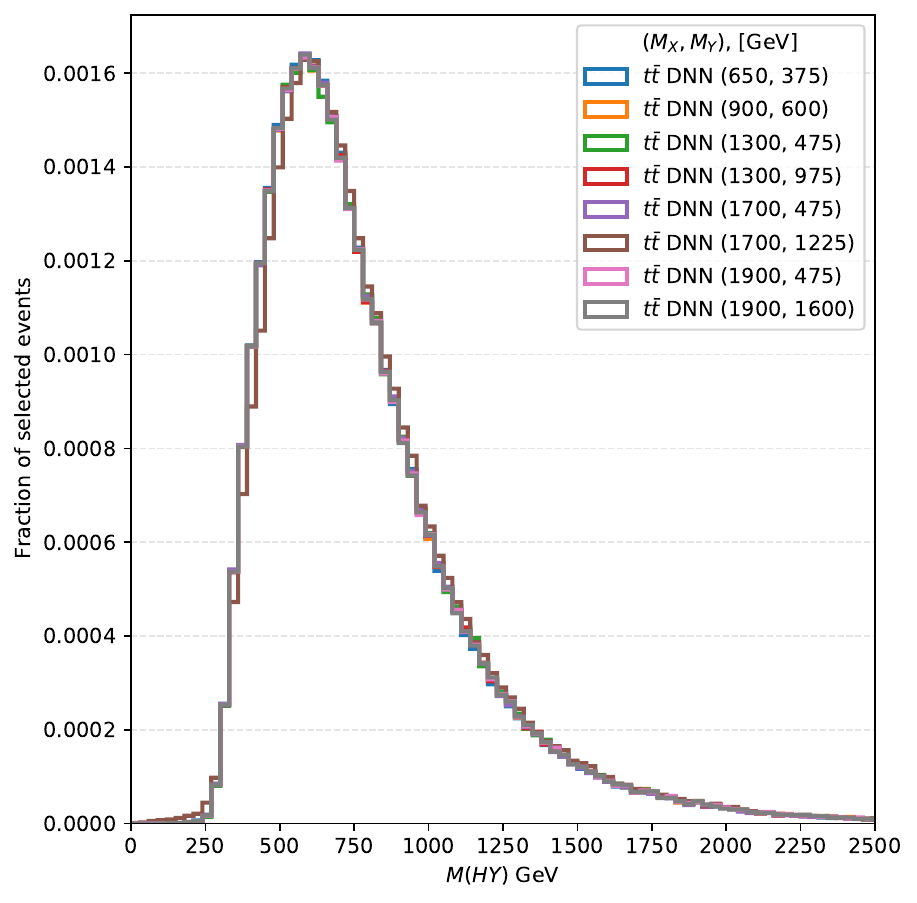} \\
  \caption{ Invariant masses reconstructed from jets selected using DNN score: pair of top-quarks candidates ($Y$ candidate, left) and Higgs and $Y$ candidates ($X$ candidate, right). Results are given for background $t\bar{t}$ process selected using DNN trained for different mass scenarios of $X$ and $Y$ resonances. }
  \label{fig:mass_combo_eval_tt}
\end{figure}

\begin{figure}
  \centering
  \includegraphics[width=0.47\columnwidth]{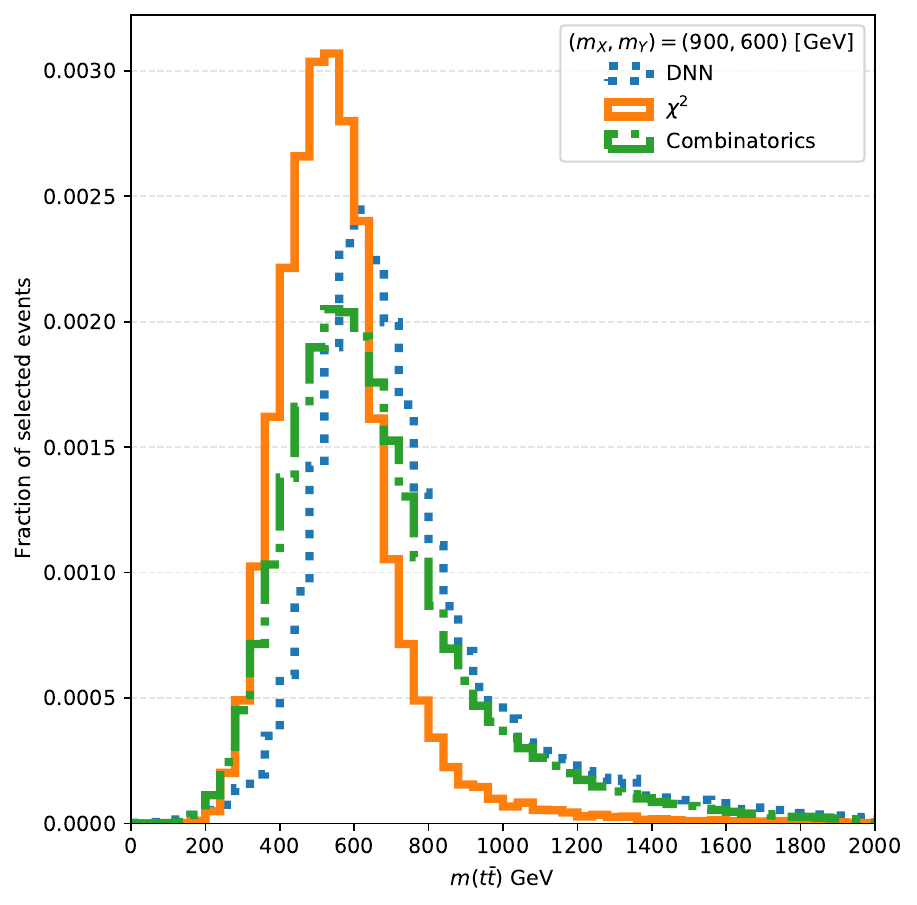}
  \includegraphics[width=0.47\columnwidth]{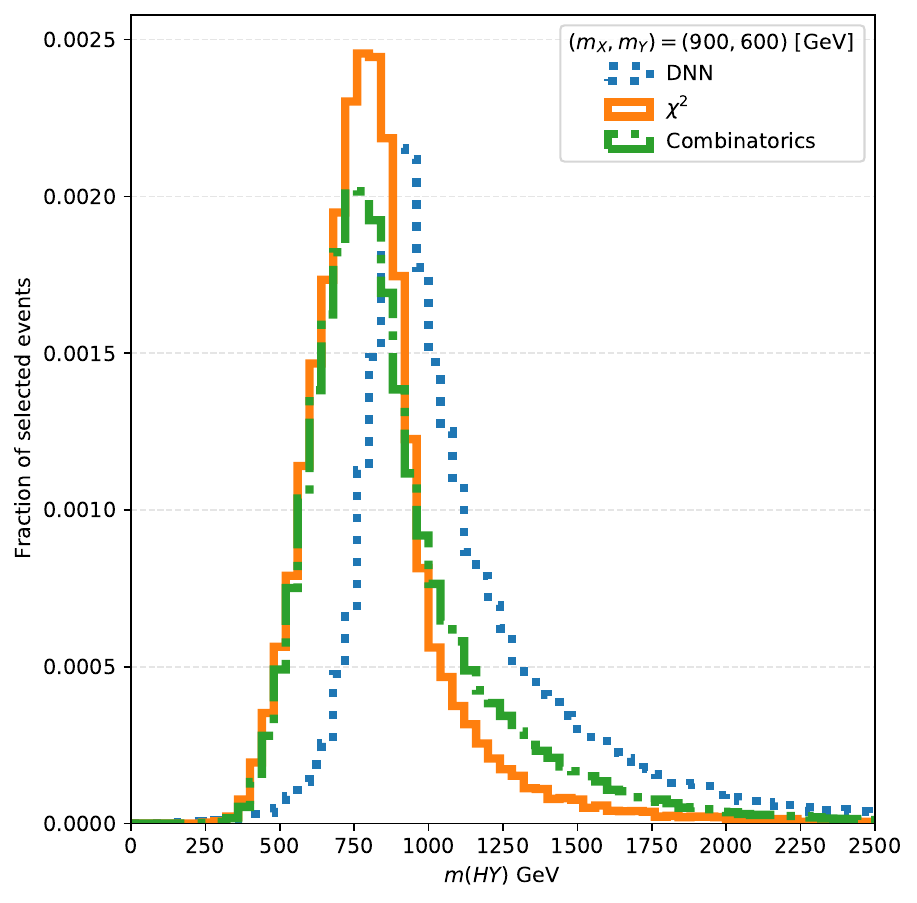} \\
  \includegraphics[width=0.47\columnwidth]{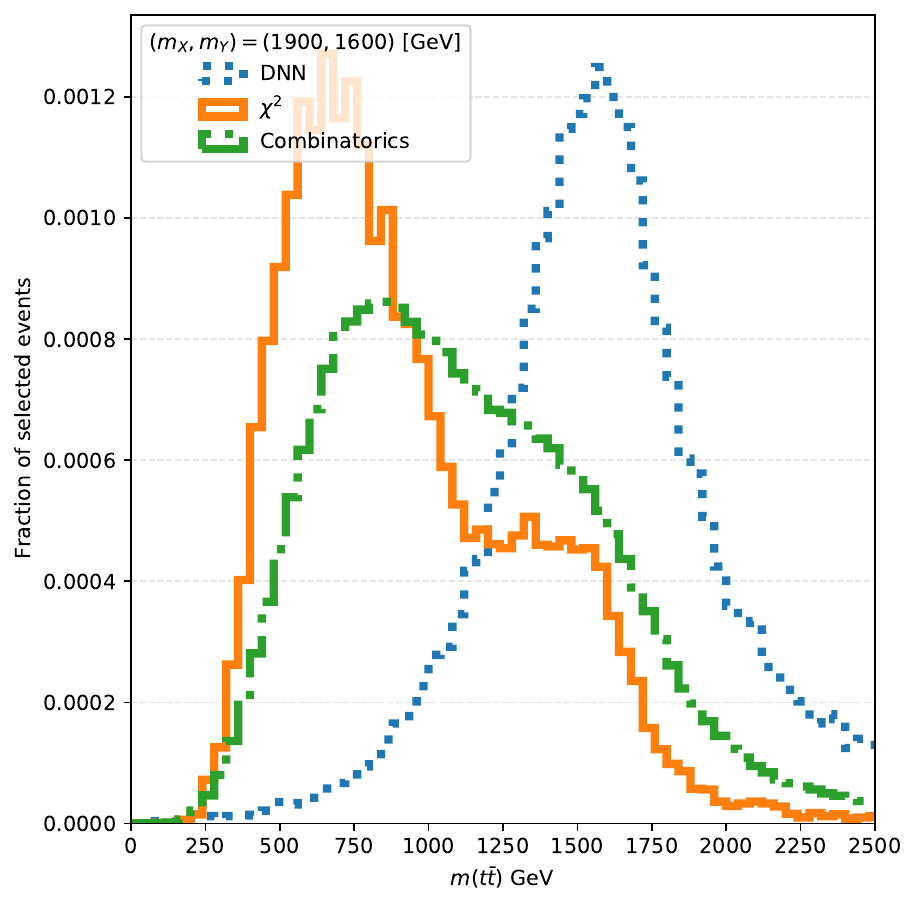}
  \includegraphics[width=0.47\columnwidth]{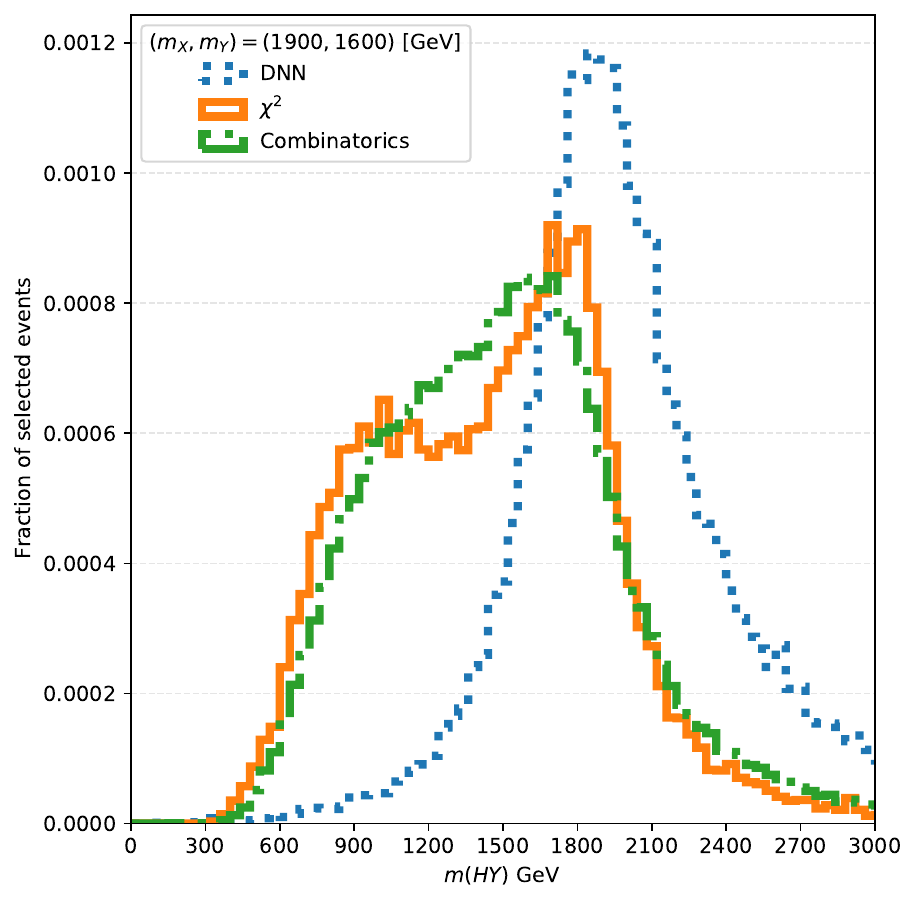} \\
  \caption{ Comparison of $X$ and $Y$ invariant masses reconstructed from all possible combinations of jets, from jets selected using $\chi^2$ metric eq. (\ref{metric}) and DNN score for $(900, 600)$ and $(1900, 1600)$ GeV mass scenarios. }
  \label{fig:comparison}
\end{figure}

Based on neural network output jets are uniquely associated with clusters with priority given to highest DNN score.
Fraction of the jets associated correctly to the truth cluster is about $60\%$ and flat over different mass points.
Fraction of events where all jets were assigned correctly is increased with the mass of the resonances from 13$\%$ for $m_X, m_Y = (900, 600)$ GeV to 26$\%$ for $m_X, m_Y = (1900, 1600)$ GeV.
In comparison, fraction of the jets associated correctly to the truth cluster using metric eq. (\ref{metric}) do not exceed $21\%$ and 
fraction of events where all jets were assigned correctly is below $3\%$.
Thus, for the scenarios with high resonance masses we found the largest DNN outperformance over jets set selection based on metric eq. (\ref{metric}) in term of $m_X$ and $m_Y$ reconstructed masses resolution (see Figure \ref{fig:mass_combo_eval} and Tab. \ref{tab:tab_reso_1}). While the distributions for background $t\bar{t}$ process found to be rather stable under different DNNs applications (see Figure \ref{fig:mass_combo_eval_tt}). The comparison of the resonance masses reconstructed using $\chi^2$ metric eq. (\ref{metric}) and DNN score is also given at Figure \ref{fig:comparison} for light and heavy mass points.

\begin{figure}
  \centering
  \includegraphics[width=0.47\columnwidth]{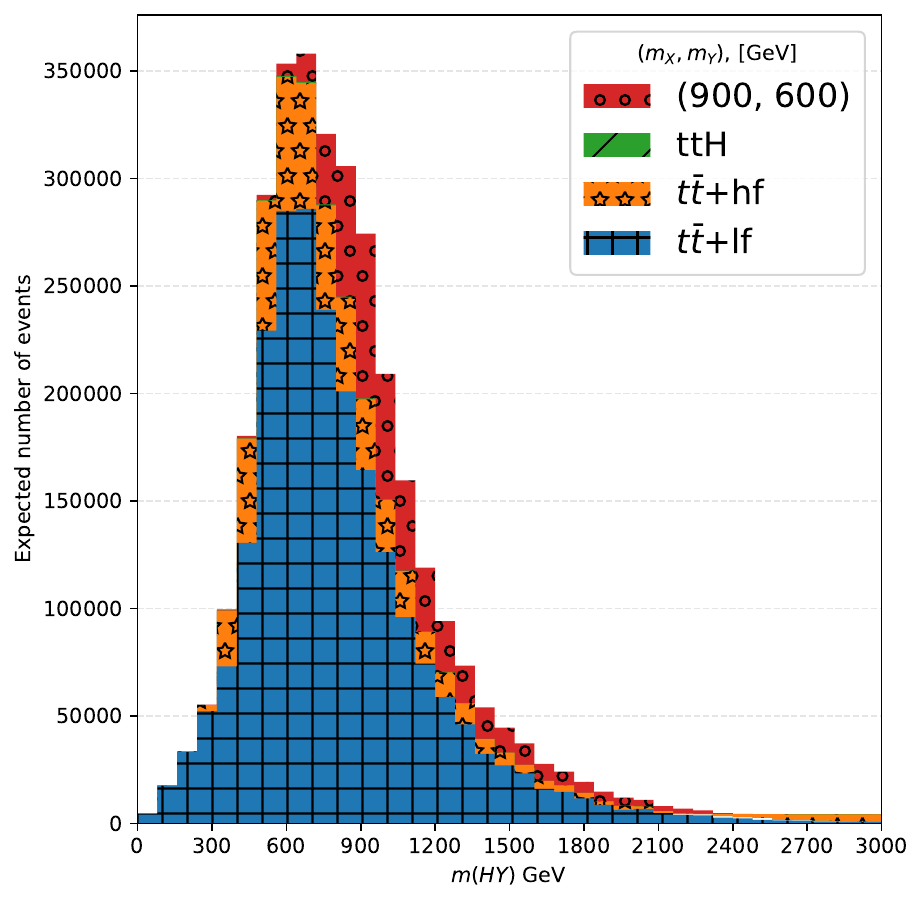} 
  \includegraphics[width=0.47\columnwidth]{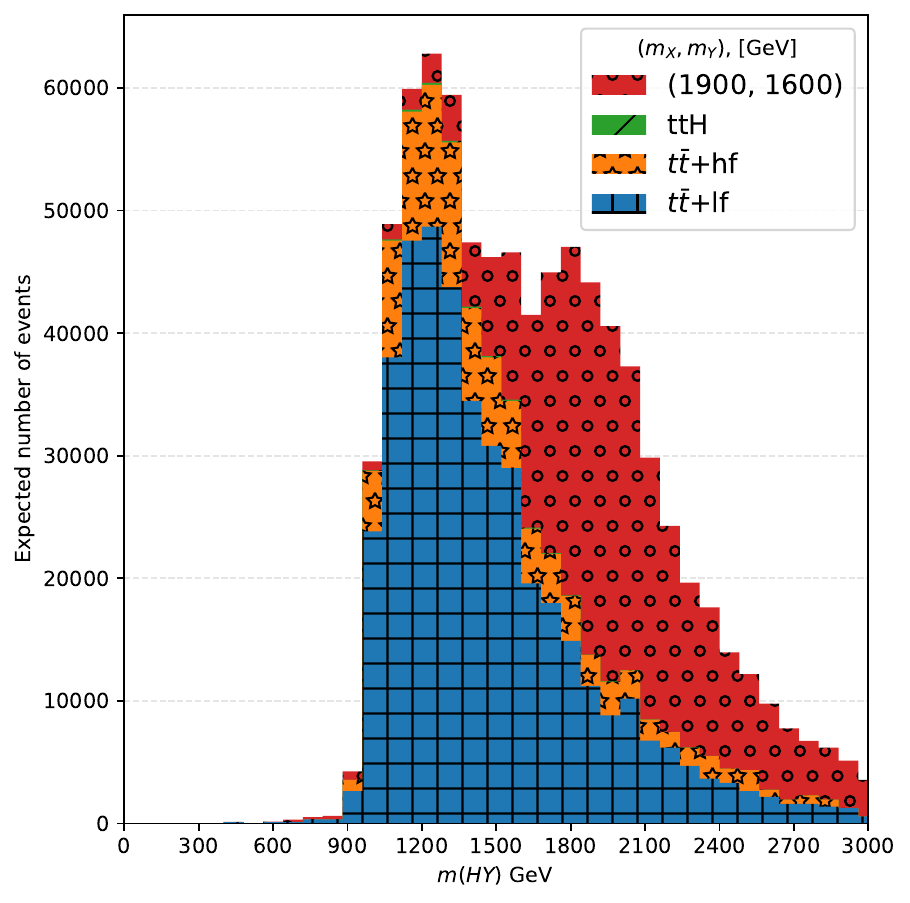}
  \caption{ Expected number of selected events at 13 TeV and integrated luminosity of 137 fb$^{-1}$ for signal and background processes in $X$ invariant mass reconstructed from jets selected using DNN score.
            Signal cross section is normalised to 2.5\% of $t\bar{t}$ cross section. }
  \label{fig:mass_combo_eval_fnal}
\end{figure}

On the top of the DNN reconstructed events we apply additional selections to suppress the backgrounds:
\begin{itemize}
  \item $m(t_{\ell}^{DNN}) > 135$ GeV
  \item $m(H^{DNN}) > 95$ GeV
  \item $m(Y^{DNN}) > 0.5 \times m_Y$ GeV
\end{itemize}
The distributions of the $m_X$ at Fig.~\ref{fig:mass_combo_eval_fnal} obtained after additional selections are used directly to perform a statistical analysis in Section \ref{section_b3}.
The expected background and signal yield and selection efficiency are given in Tab.~\ref{tab:tab_yeld_fnal} and  Tab.~\ref{tab:tab_eff_fnal}.

\begin{table}
\begin{center}
\begin{scriptsize}
\begin{tabular}{c|c|c|c|c}
$(m_{X}, m_{Y})$ & $(\overline{m}_{X}^{\chi^2} - m_{X})/m_{X}$ & $(\mu_{X}^{\chi^2} - m_{X})/m_{X}$ & $(\overline{m}_{X}^{DNN} - m_{X})/m_{X}$ & $(\mu_{X}^{DNN} - m_{X})/m_{X}$ \\ \hline
    (650, 375) &                                $-13.53 \pm 25.61$ &                              $-5 \pm 40$ &                              $10.69 \pm 33.16$ &                            $2 \pm 10$ \\
    (900, 600) &                                $-16.53 \pm 25.21$ &                             $-10 \pm 11$ &                              $11.07 \pm 31.17$ &                           $-1 \pm 10$ \\
   (1300, 475) &                                $-36.55 \pm 28.99$ &                              $-9 \pm 14$ &                              $23.23 \pm 29.76$ &                            $0 \pm 28$ \\
   (1300, 975) &                                $-20.22 \pm 27.90$ &                             $-10 \pm 12$ &                               $9.49 \pm 28.95$ &                           $-1 \pm 13$ \\
   (1700, 475) &                                $-57.15 \pm 32.54$ &                               $-5 \pm 7$ &                              $30.26 \pm 29.56$ &                            $3 \pm 11$ \\
  (1700, 1225) &                                $-25.94 \pm 30.34$ &                               $-6 \pm 6$ &                               $8.55 \pm 26.97$ &                           $-1 \pm 11$ \\
   (1900, 475) &                                $-71.36 \pm 34.45$ &                              $-5 \pm -7$ &                              $32.58 \pm 29.39$ &                            $3 \pm 14$ \\
  (1900, 1600) &                                $-27.61 \pm 34.73$ &                             $-11 \pm 14$ &                               $5.94 \pm 26.30$ &                           $-2 \pm 11$ \\
\end{tabular}
\end{scriptsize}
  \caption{ \label{tab:tab_reso_1} mismatch of $m_{X}$ $\pm$ resolution reconstructed using $\chi^2$ metric eq. (\ref{metric}) and DNN approach: $\overline{m}_{X}$ is a mean value of the distribution, $\mu_{X}$ is position of the center of the peak extracted following a Gaussian fit to the distributions.}
\end{center}
\end{table}

\begin{table}
\begin{center}
\begin{scriptsize}
\begin{tabular}{c|c|c|c}
   $(m_X, m_Y)$ &  $t\bar{t}$+lf &  $t\bar{t}$+hf &  $t\bar{t}H$ \\ \hline
   (650, 375) &        2335037 &         532262 &        4671 \\
   (900, 600) &        2293887 &         503636 &        4270 \\
  (1300, 475) &        1906047 &         395632 &        3956 \\
  (1300, 975) &        1599594 &         360848 &        3115 \\
  (1700, 475) &        1951998 &         386508 &        3969 \\
 (1700, 1225) &         994460 &         222622 &        2052 \\
  (1900, 475) &        2326007 &         498846 &        4623 \\
 (1900, 1600) &         426589 &         102757 &        1365 \\
\end{tabular}
\end{scriptsize}
  \caption{ \label{tab:tab_yeld_fnal} expected number of events after application of selections using information from DNN matched jets for different mass point (ass DNN performance and event selection are changing between benchmarks).
            Normalization is chosen to fit luminosity of 137 fb$^{-1}$ and conditions of proton-proton collisions collected by the CMS detector during Run II. }
\end{center}
\end{table}

\begin{table}
\begin{center}
\begin{scriptsize}
\begin{tabular}{c|c|c|c|c}
  $(m_X, m_Y)$ &  $t\bar{t}$+lf &  $t\bar{t}$+hf &  $t\bar{t}H$ &  Signal \\ \hline
    (650, 375) &          0.027 &          0.020 &       0.067 &   0.086 \\
    (900, 600) &          0.026 &          0.019 &       0.061 &   0.157 \\
   (1300, 475) &          0.022 &          0.015 &       0.057 &   0.138 \\
   (1300, 975) &          0.018 &          0.013 &       0.045 &   0.164 \\
   (1700, 475) &          0.022 &          0.014 &       0.057 &   0.094 \\
  (1700, 1225) &          0.011 &          0.008 &       0.030 &   0.155 \\
   (1900, 475) &          0.027 &          0.018 &       0.067 &   0.081 \\
  (1900, 1600) &          0.005 &          0.004 &       0.020 &   0.112 \\
\end{tabular}
\end{scriptsize}
  \caption{ \label{tab:tab_eff_fnal} event selection efficiency obtained using information from DNN matched jets for different mass point. }
\end{center}
\end{table}

\iffalse
Moreover, different categories based on jets assignation probability could be defined for the events, where full reconstruction is not possible and rather should not be considered.
In this study we are using following categories: 
\begin{eqnarray}
 C_{1} = & \{ \varnothing, ~l, ~\nu,~ b_{t}^{q}, ~q^{1}_{t}, ~q^{2}_{t},~ b^{1}_{H}, ~ b^{2}_{H} \} \\
 C_{2} = & \{ b_{t}^{l}, ~l, ~\nu,~ \varnothing, ~q^{1}_{t}, ~q^{2}_{t},~ b^{1}_{H}, ~ b^{2}_{H} \} \\
 C_{3} = & \{ b_{t}^{l}, ~l, ~\nu,~ b_{t}^{q}, ~q^{1}_{t}, \varnothing,~ b^{1}_{H}, ~ b^{2}_{H}  \}  \\
 C_{4} = & \{ b_{t}^{l}, ~l, ~\nu,~ b_{t}^{q}, ~q^{1}_{t}, ~q^{2}_{t},~ b^{1}_{H}, ~ \varnothing \} \\
 C_{\textrm{other}} = & \{ ...  \}
\label{rat4}
\end{eqnarray}
where $C_{1}$ is a category for events without reconstruction of b-tagged jet from leptonic top-quark decay,
$C_{2}$ - without b-tagged jet from hadronic top-quark decay,
$C_{3}$ - without one light jets from  hadronic top-quark decay,
$C_{4}$ - without one b-tagged jets from  $H$ decay.
All events outside $C_{\textrm{full}}$ and $C_{1} - C_{4}$ categories fall into $C_{\textrm{other}}$ category.
\fi

\subsection{ Statistical analysis and results } \label{section_b4}
Frequentist inference is performed using CombinedLimit package \cite{CMS-NOTE-2011-005} to extract expected exclusion limits at 95\% C.L based on $X$ invariant mass binned distribution with a good separation
of signal and background events. 
Overall pre-fit uncertainty on $t\bar{t}H$ and $t\bar{t}$ backgrounds in CMS $t\bar{t}H$ searches \cite{CMS:2018hnq, CMS-PAS-HIG-17-026} in different per jet-process categories were estimated to not exceed $25\%$
with largest contributions from the theoretical uncertainties.
Thus, for the SM $t\bar{t}-$hf background a 50\% normalization uncertainty is introduced following \cite{CMS:2018hnq} and for SM $t\bar{t}-$lf and $t\bar{t}H$ a conservative 30\% normalization uncertainty is incorporated in statistical model as nuisance parameter.
The cross section of the $t\bar{t}$ for $pp$ collisions at a center-of-mass energy of $\sqrt{s} = 13$ TeV is $831\pm{51}$~pb ($985.7$~pb at $\sqrt{s} = 14$ TeV) for a top quark mass of 172.5 GeV \cite{Czakon:2011xx, Botje:2011sn}.
The datasets normalization correspond to an integrated luminosity of 137 fb$^{-1}$ of proton-proton collisions collected by the CMS detector during Run II \cite{CMS:2021yci}.
The asymptotic frequentist formula \cite{Cowan:2010js} is used to obtain an expected upper limit on signal cross section based on an Asimov data set of background-only model.

In addition, the reconstruction efficiency estimated in section 4 can be used to project
the resonant production searches into HL-LHC conditions, defined by total integrated luminosity of 3 ab$^{-1}$ and collision energy of 14 TeV.
For this we rescale $m_X$ background shapes using cross sections and luminosity HL-LHC to LHC ratios.

\begin{table}
\begin{center}
\begin{scriptsize}
\begin{tabular}{c|c|c|c|c}
& \multicolumn{3}{c}{ 13 TeV, 137 fb$^{-1}$ } & 14 TeV, 3000 fb$^{-1}$ \\
 $(m_X, m_Y)$ [GeV] & Expected U.L.       &      TRSM &                     NMSSM  & Expected U.L.  \\  \hline
   (650, 375) &         3278 &       5.9 &                9.3 at (600, 400)  & 170 \\
   (900, 600) &         804 &       1.1 &                2.5 at (900, 600)  & 44 \\
  (1300, 475) &         331 &     0.07  &              0.6 at (1200, 500)   & 17 \\
  (1300, 975) &         349 &    0.02   &  0.6 at (1200,800)              & 18 \\
  (1700, 475) &         262 &    0.006  &             0.04 at (1600, 500)  & 13 \\
 (1700, 1225) &         223 &   0.0015  &                               -  & 11 \\
  (1900, 475) &         265 &   0.002   &             0.01 at (1800, 500)   & 13 \\
 (1900, 1600) &         223 &  0.0002   &                               -  & 11 \\ 
\end{tabular}
\end{scriptsize}
  \caption{ \label{tab:tab_resultsA} expected 95\% upper limits [fb] on $\sigma(X)\times Br(X\rightarrow HY) \times Br(Y \rightarrow t\bar{t})$ at $95\%$ CL at LHC Run II and HL-LHC conditions. 
  Comparison with TRSM and NMSSM \cite{Ellwanger:2022jtd} predictions for 13 TeV are shown for the closest available mass points.
  }
\end{center}
\end{table}

\iffalse
\begin{table}
\begin{center}
\begin{scriptsize}
\begin{tabular}{c|c|c|c}
 $(m_X, m_Y)$ [GeV] & Expected Limits       &      CMS \cite{CMS:2022suh} & CMS \cite{CMS:2021yci}  \\  \hline
   (650, 375) &                     3278 &                                         - &    335.3 at (600, 350) \\
   (900, 600) &                     804 &                                            - &    109.1 at (900, 600) \\
  (1300, 475) &                     331 &      116.5 at (1300, 450)              &    38.9 at (1400, 450) \\
  (1300, 975) &                     349 &                                         - &   46.1  at (1400, 1000) \\
  (1700, 475) &                     262 &      24.9 at (1700, 450)               &    82.0 at (1800, 450) \\
 (1700, 1225) &                     223 &                                        - &  31.2 at (1800, 1200) \\
  (1900, 475) &                     265 &     11.4 at (1900, 450)               &   117.4 at (1900, 450) \\
 (1900, 1600) &                     223 &                                      - &   34.5 at (1900, 1600) \\
\end{tabular}
\end{scriptsize}
  \caption{ \label{tab:tab_resultsB} expected upper limits [fb] on $\sigma(X)\times Br(X\rightarrow HY) \times Br(Y \rightarrow t\bar{t})$ in comparison with CMS Run II measurements
   of similar processes in $Y \rightarrow b\bar{b}$ decay channel. 
  The expected limits correspond to an integrated luminosity of 137 fb$^{-1}$ of proton-proton collisions. CMS observations are shown for the closest available mass points. }
\end{center}
\end{table}
\fi

\begin{table}
\begin{center}
\begin{scriptsize}
\begin{tabular}{c|c|c|c}
 $(m_X, m_Y)$ [GeV] & Expected Limits    &      CMS \cite{CMS:2022suh} & CMS \cite{CMS:2021yci}  \\  \hline
   (650, 375) &                     1909 &                                         - &    335.3 at (600, 350) \\
   (900, 600) &                     468  &                                            - &    109.1 at (900, 600) \\
  (1300, 475) &                     193  &      116.5 at (1300, 450)              &    38.9 at (1400, 450) \\
  (1300, 975) &                     203  &                                         - &   46.1  at (1400, 1000) \\
  (1700, 475) &                     152  &      24.9 at (1700, 450)               &    82.0 at (1800, 450) \\
 (1700, 1225) &                     130  &                                        - &  31.2 at (1800, 1200) \\
  (1900, 475) &                     154  &     11.4 at (1900, 450)               &   117.4 at (1900, 450) \\
 (1900, 1600) &                     130  &                                      - &   34.5 at (1900, 1600) \\
\end{tabular}
\end{scriptsize}
  \caption{ \label{tab:tab_resultsB} expected upper limits [fb] on $\sigma(X) \times Br(X\rightarrow HY) \times Br(Y \rightarrow t\bar{t}) \times Br(H \rightarrow b\bar{b})$ in comparison with CMS Run II measurements
   of processes with $X \rightarrow YH$, $Y \rightarrow bb$, $H \rightarrow bb$ \cite{CMS:2022suh} and $X \rightarrow YH$, $Y \rightarrow bb$, $H \rightarrow \tau\tau$ \cite{CMS:2021yci} decay chains.
  The expected limits correspond to an integrated luminosity of 137 fb$^{-1}$ of proton-proton collisions. CMS observations are shown for the closest available mass points. }
\end{center}
\end{table}

The results of the statistical analysis based on histograms obtained using DNN analysis strategy (Section \ref{section_b3}) are given at Table \ref{tab:tab_resultsA},
where possible cross sections at 13 TeV for $ggF \rightarrow X \rightarrow (Y \rightarrow t\bar{t}) + H$ given by NMSSM are also shown.
The NMSSM cross sections satisfied broad range of limitation from the existing searches, constraints from theoretical and experimental sources \cite{Ellwanger:2022jtd}.
Predictions of TRSM for production and decay rates are obtained from ScannerS \cite{Coimbra:2013qq, Muhlleitner:2020wwk}, used to perform a flat scans in TRSM parameter space of
dimension seven. 
For this scan in $t\bar{t}b\bar{b}$ final state we repeat procedure described in \cite{Tania}.
Large number of theoretical constraints described in \cite{Robens:2019kga} and experimental constraints implemented in {\scshape HiggsBounds } \& {\scshape HiggsSignals } \cite{Bechtle:2008jh, Bechtle:2013xfa} tools are taken into account.
No remaining parameter space of TRSM or NMSSM models can be probed with expected sensitivity in considered mass points.

CMS Run II measurements of 
$\sigma(X) \times Br(X\rightarrow HY) \times Br(Y \rightarrow b\bar{b}) \times Br(H \rightarrow b\bar{b})$ and
$\sigma(X) \times Br(X\rightarrow HY) \times Br(Y \rightarrow b\bar{b}) \times Br(H \rightarrow \tau \bar{\tau})$
are given at Table \ref{tab:tab_resultsB} for indirect comparison with $t\bar{t}b\bar{b}$ final state results. 
For example, under TRSM in some points of the model free parameters space 
$Br(Y \rightarrow t\bar{t})$ can be 3000 times greater than $Br(Y \rightarrow b\bar{b})$ at $(1300, 975)$ mass point. 
In this case the overall $pp \rightarrow X \rightarrow Y(b\bar{b})H$ decay channel will have better sensitivity to $X$ boson production.

\section{ Conclusions } \label{section_c}
A probe of a search for the decay of a heavy scalar boson $X$ into the observed Higgs boson $H$ and another scalar boson $Y$ has been presented. 
The $H$ and the $Y$ bosons are required to decay into a pair of $b$ quarks and a pair of top quarks, respectively. Semileptonic decay of top quarks is considered.
The search is projected on operation conditions of CMS detector during the LHC Run II data taking period at a center-of-mass energy of 13 TeV.
Realistic objects and events selections are applied, allowing to suppress most of the backgrounds.
Machine learning approach using Deep Neural Network is proposed to resolve ambiguous in jets assignment and improve kinematic reconstruction of signal events.
Detector sensitivity is obtained as $95\%$ expected upper limits on the product of the production cross section and the branching fractions of the searched anomalous process.
The outcome of our study is summarized at Table \ref{tab:tab_resultsA} showing limits in the range from 3278 fb $m_X = 650$ GeV to 223 fb for $m_X = 1900$ GeV. 
Further improvements could be achieved through the combination of searches results of different top quarks and Higgs boson decays channels.
The proposed searches strategy could be considered as a road map for real data analysis at the LHC.

\section*{Acknowledgments}
We  would like to thank S.~Slabospitskii for useful discussions. This work is supported by the Russian Science Foundation under grant 21-72-00098.

\label{app_benchmarks}

\bibliographystyle{elsarticle-num} 
\biboptions{numbers,sort&compress}
\bibliography{XYZ_1.bib}


%\begin{thebibliography}{00}

% \end{thebibliography}
\end{document}